\RequirePackage{fix-cm}
\documentclass[smallextended]{svjour3}           % onecolumn (second format)
\usepackage{graphicx}
\usepackage{amssymb}
\usepackage{amsmath}
\usepackage{multirow}
\usepackage{subfigure}
\usepackage{color}
\usepackage{setspace}
\journalname{Experimental Astronomy}
\begin{document}

\title{The design and flight performance of the PoGOLite Pathfinder balloon-borne hard X-ray polarimeter}

\titlerunning{Design and flight performance of the PoGOLite Pathfinder}        % if too long for running head

\author{M. Chauvin \and  
H.-G. Flor\'en  \and 
M. Jackson  \and 
T. Kamae  \and 
T. Kawano  \and 
M. Kiss \and 
M. Kole \and 
V. Mikhalev  \and 
E. Moretti \and 
G. Olofsson \and 
S. Rydstr\"{o}m \and 
H. Takahashi \and 
J. Lind  \and 
J.-E. Str\"{o}mberg \and 
O. Welin \and 
A. Iyudin  \and 
D. Shifrin \and
M. Pearce
}

%\institute{F. Author \at
%             first address \\
%             Tel.: +123-45-678910\\
%              Fax: +123-45-678910\\
%            \email{fauthor@example.com}           %  \\
%          S. Author \at
%             second address
%}

\institute{M. Chauvin, M. Jackson, M. Kiss, M. Kole, V. Mikhalev, E. Moretti, S. Rydstr\"{o}m, M. Pearce 
\at
KTH Royal Institute of Technology, Department of Physics, SE-106~91~Stockholm, Sweden, 
and The Oskar Klein Centre for Cosmoparticle Physics, AlbaNova University Centre, 106~91~Stockholm, Sweden.  \\
\email{pearce@kth.se, mozsi@kth.se} \\
\emph{Present address} of M. Kole: Geneva University, CH-1211 Geneva, Switzerland. \\
\emph{Present address} of E. Moretti: Max-Planck-Institut f\"ur Physik, D-80805 Munich, Germany. 
%              \email{fauthor@example.com}           %  \\
     \and
     H.-G. Flor\'en, G. Olofsson \at
     Stockholm University, Department of Astronomy, AlbaNova University Centre, SE-106~91~Stockholm, Sweden. 
	\and
	T. Kamae \at
	University of Tokyo, Department of Physics, 113-0033 Tokyo, Japan.
	\and
	T. Kawano, H. Takahashi \at
	Hiroshima University, Department of Physical Science, 739-8526 Hiroshima, Japan. 
	\and
	J. Lind, J.-E. Str\"{o}mberg, O. Welin \at	
	DST Control, {\AA}kerbogatan 10, 582~54~Link\"{o}ping, Sweden.
	\and
	A. Iyudin \at
	Skobeltsyn Institute of Nuclear Physics, Moscow State University by M. V. Lomonosov, 119991 Moscow, Russia.
	\and
	D. Shifrin \at
	Russian Federal Service for Hydrometeorology and Environmental Monitoring, Central Aerological Observatory, Moscow, Russia.
	}
\date{Received: date / Accepted: date}

\maketitle

%\doublespacing

%%%%%%%%%%%%%%%%%%%%%%%%%%%%%%%%%%%%%%%%%%%%%%%%%%%%%%%%%%%%%%%%%%%
\begin{abstract}

In the 50 years since the advent of X-ray astronomy there have been many scientific advances due to the development of new experimental 
techniques for detecting and characterising X-rays. Observations of X-ray polarisation have, however, not undergone a similar development. 
This is a shortcoming since a plethora of open questions related to the nature of X-ray sources could be resolved through measurements 
of the linear polarisation of emitted X-rays. 
The PoGOLite Pathfinder is a balloon-borne hard X-ray polarimeter operating in the \mbox{25 - 240 keV} energy band from a stabilised 
observation platform.
Polarisation is determined using coincident energy deposits in a segmented array of plastic scintillators surrounded by a BGO 
anticoincidence system and a polyethylene neutron shield. The PoGOLite Pathfinder was launched from the SSC Esrange Space Centre in 
July 2013. A near-circumpolar flight was achieved with a duration of approximately two weeks.
The flight performance of the Pathfinder design is discussed for the three Crab observations conducted. The signal-to-background 
ratio for the observations is shown to be 0.25$\pm$0.03 and the Minimum Detectable Polarisation (99\% C.L.) is (28.4$\pm$2.2)\%.
A strategy for the continuation of the PoGOLite programme is outlined based on experience gained during the 2013 maiden flight.

\end{abstract}
%%%%%%%%%%%%%%%%%%%%%%%%%%%%%%%%%%%%%%%%%%%%%%%%%%%%%%%%%%%%%%%%%%%
%
%%%%%%%%%%%%%%%%%%%%%%%%%%%%%%%%%%%%%%%%%%%%%%%%%%%%%%%%%%%%%%%%%%%
%
\section{Introduction}
\label{Introduction}
X-ray sources are known to include non-thermal radiation emitted either from asymmetric systems (e.g. jets, 
columns, and accretion disks), or from systems having ordered magnetic fields. For such sources, the emission in X-rays is 
expected to be polarised. X-ray polarimetry remains, however, a largely unexplored field of high-energy astrophysics.
Imaging, spectroscopy and timing are the standard observation techniques. The addition of the polarisation fraction and angle 
provides unique information regarding the emission mechanism and the geometry of the emitting region, respectively~\cite{background}. 

A concept for a balloon-borne hard X-ray Compton polarimeter with sensitivity for point sources with X-ray fluxes of order 100 mCrab 
has been presented previously~\cite{Overview paper}. The collecting area of the polarimeter is defined by 217 close-packed hexagonal 
cross-section plastic scintillator detector cells. 
In order to initiate the experimental programme, a reduced area polarimeter comprising 61 
detector cells has been constructed to allow observations of 1 Crab sources. 
The PoGOLite Pathfinder mission was launched from the SSC Esrange Space Centre on July 12th 2013. 
The aims of this flight were to evaluate the experimental design, study high latitude stratospheric background conditions and make 
observations of the Crab \mbox{(25 - 240~keV)}. 
Planned polarimetric observations of Cygnus X-1 were not possible since the required hard state~\cite{magnus} was not present.   

For the synchrotron processes expected in the Crab, the electric field vector of the X-ray flux will be perpendicular to the magnetic field 
lines in the emitting region. A polarisation measurement therefore determines the direction of the magnetic field allowing competing 
emission models to be tested in the hard X-ray regime. For accretion disks around black holes, such as Cygnus X-1, Compton scattering 
processes are present.
In this case, the electric vector is perpendicular to the plane of scattering and a polarisation measurement determines the geometrical 
relation between the photon source and the scatterer. The linear polarisation of X-ray emissions (2.6~keV and 5.2~keV) from the Crab 
nebula were studied over forty years ago using a dedicated polarimeter~\cite{MW}. The relatively high polarisation fraction 
observed ($\sim$19\%) is compatible with synchrotron emission. The polarisation angle was found to lie approximately 30$^\circ$ from the 
Crab pulsar spin axis.
Polarimetric observations made using instruments on-board the INTEGRAL observatory have been made more recently~\cite{integral} for 
the energy range of 100's~keV to 1~MeV. 
The polarimetric response of these instruments was not evaluated prior to launch.
The high polarisation fraction observed supports synchrotron emission from a well-ordered magnetic environment. The polarisation angle 
was determined to be well aligned to the Crab pulsar spin axis, indicating that high-energy electrons are generated close to the pulsar.  
%
%%%%%%%%%%%%%%%%%%%%%%%%%%%%%%%%%%%%%%%%%%%%%%%%%%%%%%%%%%%%%%%%%%%
%
\section{Polarimeter design}
\label{Polarimeter design}
\subsection{Measurement principle}

\begin{figure}[!hbt]
\centering
\includegraphics[width=\textwidth]{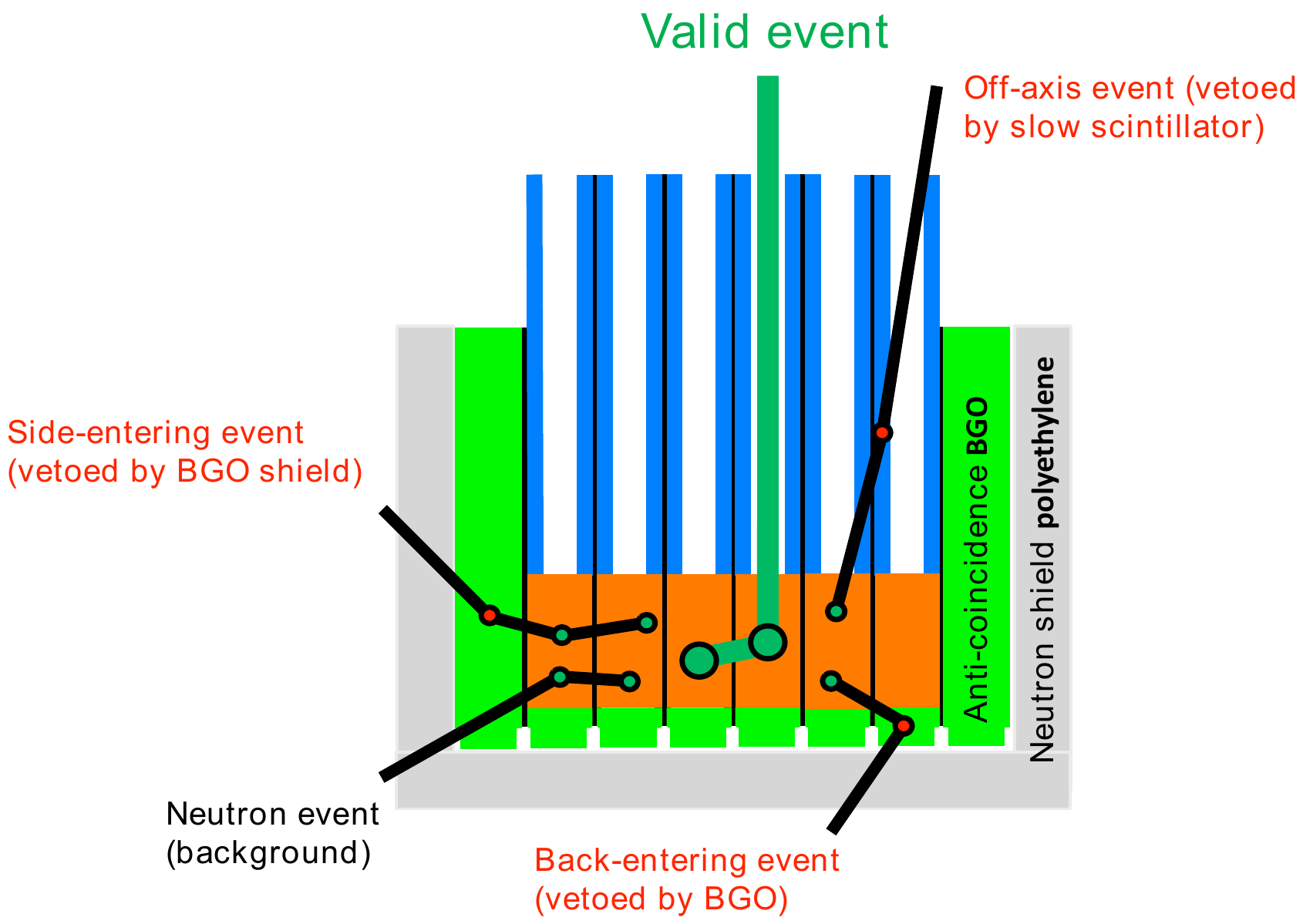}
\caption{\label{PoGOLite array sketch}
A schematic description of the polarimeter design concept.
Source X-rays pass cleanly through a plastic scintillator and metal foil collimator (blue) and may scatter between
solid plastic scintillators (orange). X-rays which impinge on the array of solid plastic scintillators from outside the field-of-view 
defined by the collimator system (2.4$^{\circ} \times 2.6^{\circ}$) and lateral BGO anticoincidence (green) will be absorbed and/or produce anticoincidence signals.
The bottom BGO scintillators (green) identify albedo charged particles or photons. A polyethylene neutron shield surrounds the scintillator array.}
\end{figure}

The PoGOLite polarimeter comprises a close-packed hexagonal array of plastic scintillators and a mixed plastic/BGO ({Bi$_{4}$Ge$_{3}$O$_{12}$})
anticoincidence system. 
The polarisation of X-rays incident on the array of plastic scintillators is determined by reconstructing the distribution of 
azimuthal scattering angles defined relative to the polarisation plane.  A polarisation event can be characterised by two energy deposits in 
the scintillator array -- corresponding to a Compton scatter and either a photoelectric absorption or a second Compton scatter in a 
nearby scintillator element.
Figure~\ref{PoGOLite array sketch} describes the measurement concept.
During observations the polarimeter is rotated (0.2 r.p.m.) around the viewing axis in order to remove the systematic bias which could 
result e.g. from intrinsic differences in efficiency between the scintillator elements. Each polarisation event is tagged with the roll angle, 
generating a continuous distribution of scattering angles. 
The resulting distribution of azimuthal scattering angles will be modulated for a polarised incident flux. 
The modulation factor is defined as the ratio between the amplitude and mean value of a fitted sinusoidal modulation curve.
The polarisation fraction of the source emission is given by the measured modulation factor divided by that for a 100\% polarised beam (M$_{100}$), derived
from computer simulations validated with tests using polarised photons~\cite{Ground calibration paper}. 
The polarisation angle is related to the phase of the fitted modulation curve. Alternatively, an 
approach based on Stokes vectors can be used to extract polarisation parameters~\cite{Stokes}.
The design of the polarimeter has been optimised and characterised using Geant4 simulations~\cite{Overview paper}, tests using polarised 
synchrotron beams~\cite{synchro} and radioactive sources~\cite{Ground calibration paper}. The expected M$_{100}$ for Crab 
observations is $\sim$22\%~\cite{Ground calibration paper}. In the remainder of this section, the implementation of the conceptual design outlined 
in~\cite{Overview paper} is described in detail.

%%%%%%%%%%%%%%%%%%%%%%%%%%%%%%%%%%%%%%%%%%%%%%%%%%%%%%%%%%%%%%%%%%%
%
\subsection{Scintillator array}
\label{Detector array design}
\begin{figure}[!hbt]
\centering
\includegraphics[width=\textwidth]{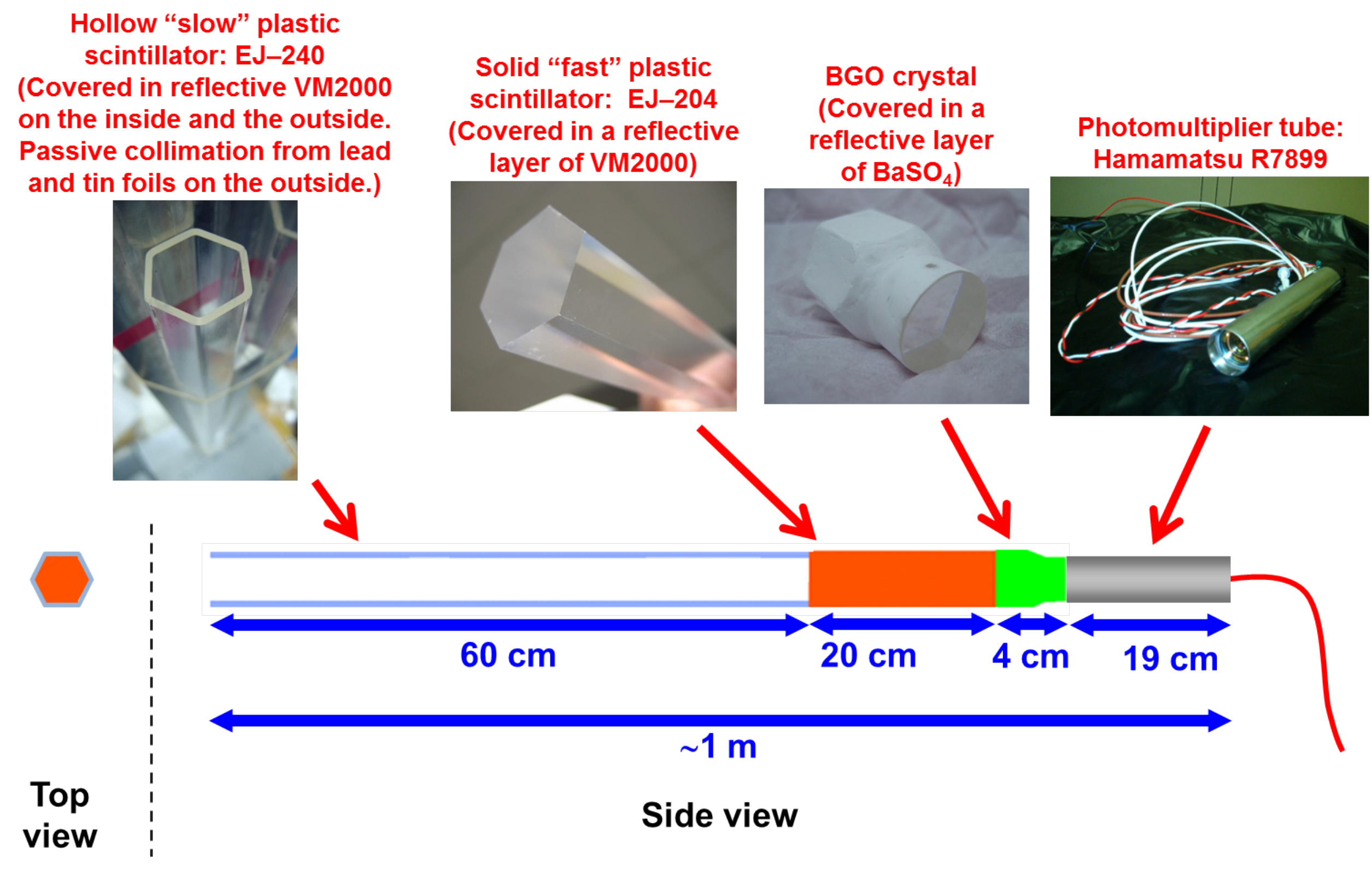}
\caption{\label{PDC components}The scintillator components of a phoswich detector cell (PDC). The width of a cell is approximately 
\mbox{2.8 cm} (side to side). The bottom BGO piece (shown in green) is tapered and provides a transition from the hexagonal cross-
section of the fast scintillator (red) to the round photomultiplier tube window.}
\end{figure}

The scintillator array comprises 61 hexagonal cross-section phoswich detector cells (PDCs) arranged in a honeycomb structure. Each PDC 
is \mbox{84 cm} long and has three scintillating elements, as shown in Figure~\ref{PDC components}. 
The upper part of the PDC is an active collimator ``slow'' plastic scintillator tube (Eljen Technology EJ$-$240, \mbox{60 cm} long, \mbox{2 mm} wall thickness, 
scintillation decay time 285~ns), while the bottom part is a anticoincidence BGO crystal (\mbox{4 cm} long, 
scintillation decay time 300~ns, supplied by the Nikolaev Institute of Inorganic Chemistry). Sandwiched between these two anticoincidence 
components is a solid ``fast'' plastic scintillator rod (EJ$-$204, \mbox{20 cm} long, scintillation decay time 1.8~ns). 
The scintillators are glued together using a UV transparent polyurethane-based epoxy resin.
The fast scintillator rods and slow scintillator tubes  are wrapped (inside and outside) in a reflective layer of 80~$\mu$m thick 3M VM2000 
radiant mirror film to maximise the light collection. Due to their irregular shape, the BGO crystals are instead covered in a layer of $\sim$350~$
\mu$m thick epoxy loaded with reflective BaSO$_4$. 
For additional passive collimation, each slow scintillator tube is sheathed in \mbox{50 $\mu$m} thick foils of lead and tin (to absorb 
fluorescence X-rays from the lead). The fast and slow scintillators are covered in a 200~$\mu$m thick protective layer of plastic heat-shrink film. Each PDC 
is read out by a modified version of the Hamamatsu Photonics R7899 photomultiplier tube (PMT). 
The PMT is designed to have a sufficiently low dark current level to allow Compton energy 
deposits from incident \mbox{25 keV} photons ($\sim$1.2~keV corresponding to $\sim$0.6 photo-electrons is deposited for a polar 
scattering angle of 90$^\circ$) to be detected. A \mbox{1 mm} thick silicone wafer, Eljen Technology \mbox{EJ$-$560}, and generic optical 
grease are used to interface the BGO crystal surface to the photomultiplier tube window. Each PDC is co-aligned with the polarimeter 
viewing axis with a precision exceeding \mbox{$\sim$0.1$^{\circ}$}.
The plastic scintillator array is surrounded by a \mbox{60 cm} tall BGO side anticoincidence shield (SAS) comprising 30 units, where each 
unit consists of three \mbox{20 cm} long segments, glued together, and mounted on a 10~mm thick aluminium support. 
SAS BGO crystals are covered in a reflective layer of BaSO$_4$ and read out using the same PMT as used for the PDCs. 
 
\begin{figure}[!hbt]
\centering
\includegraphics[width=.45\textwidth]{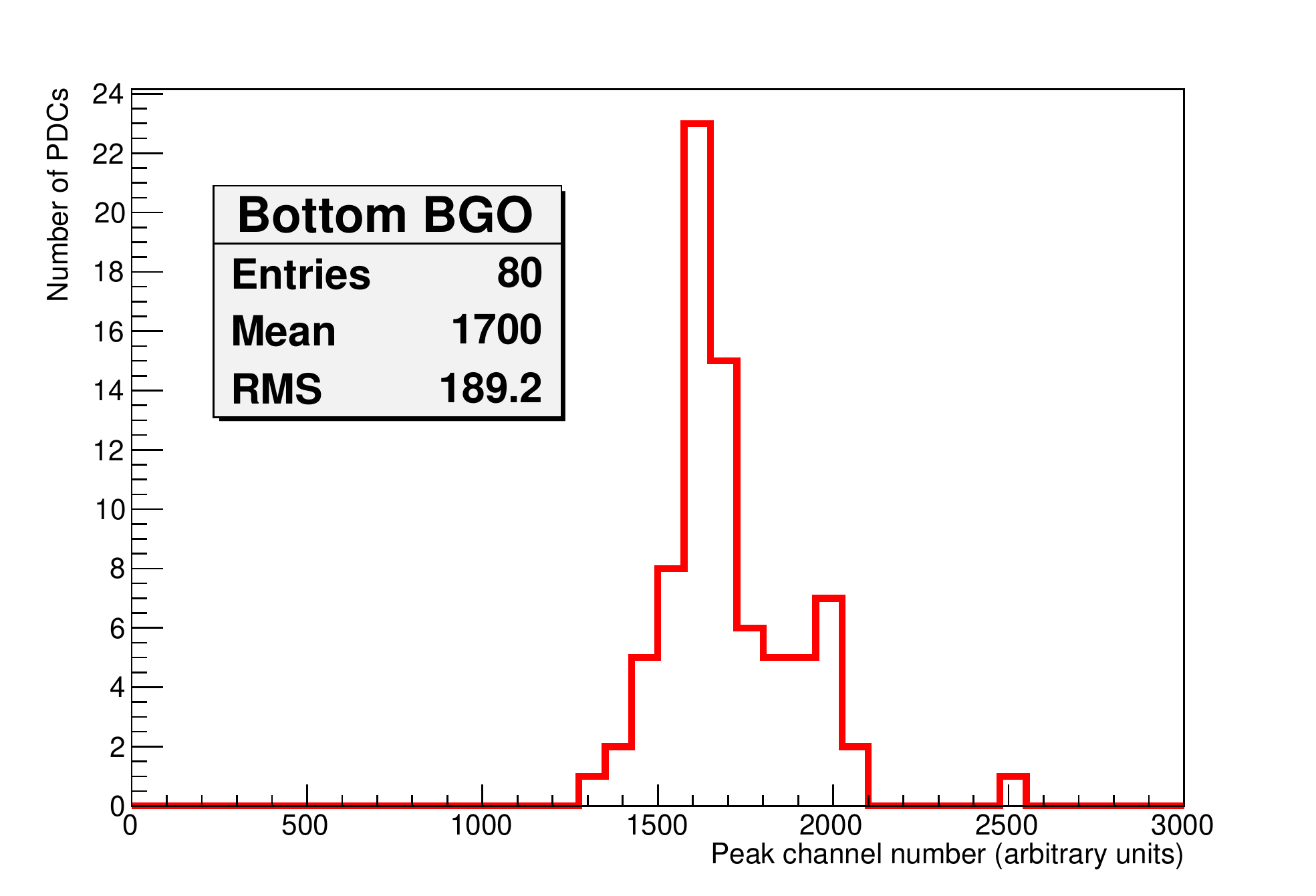}
\includegraphics[width=.45\textwidth]{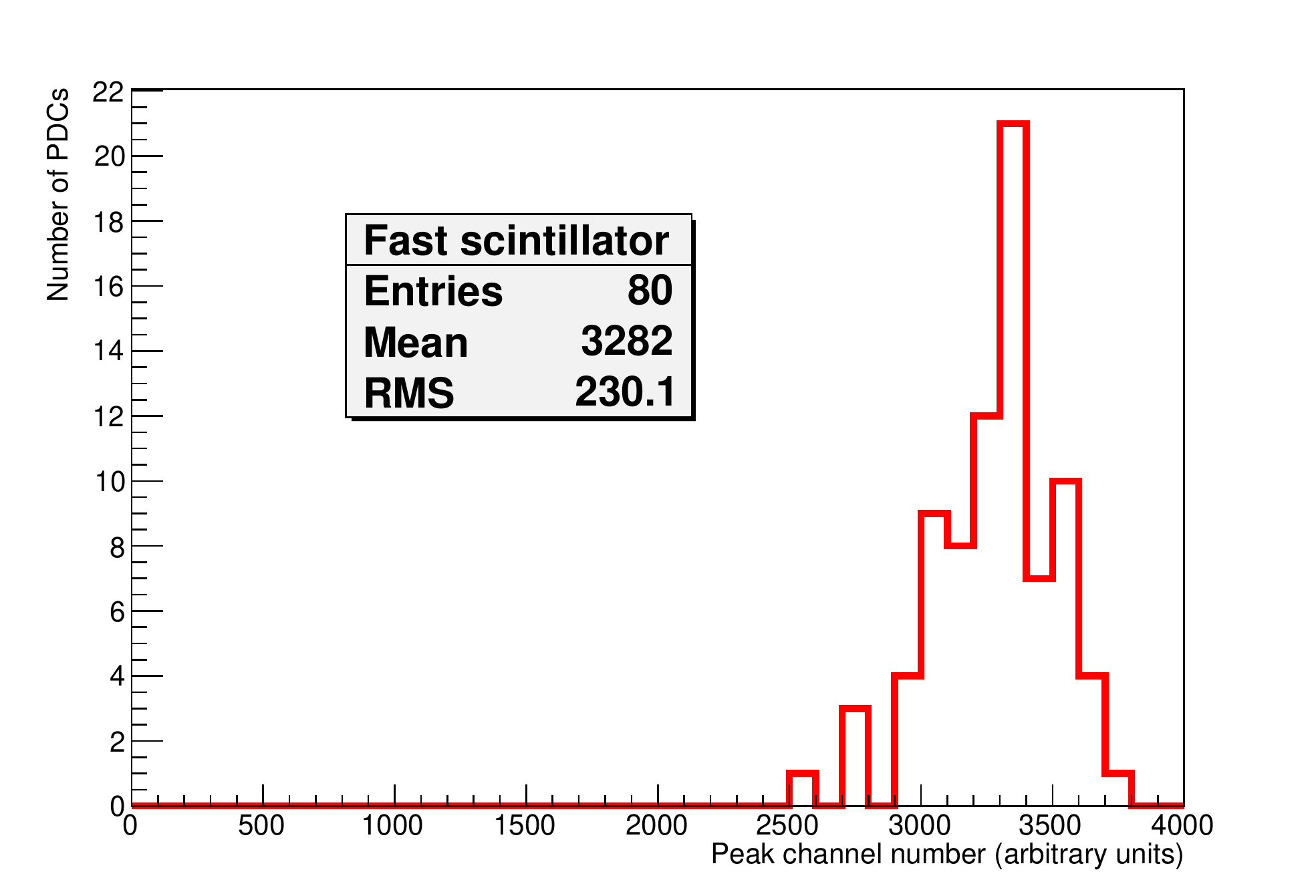} 
\includegraphics[width=.75\textwidth]{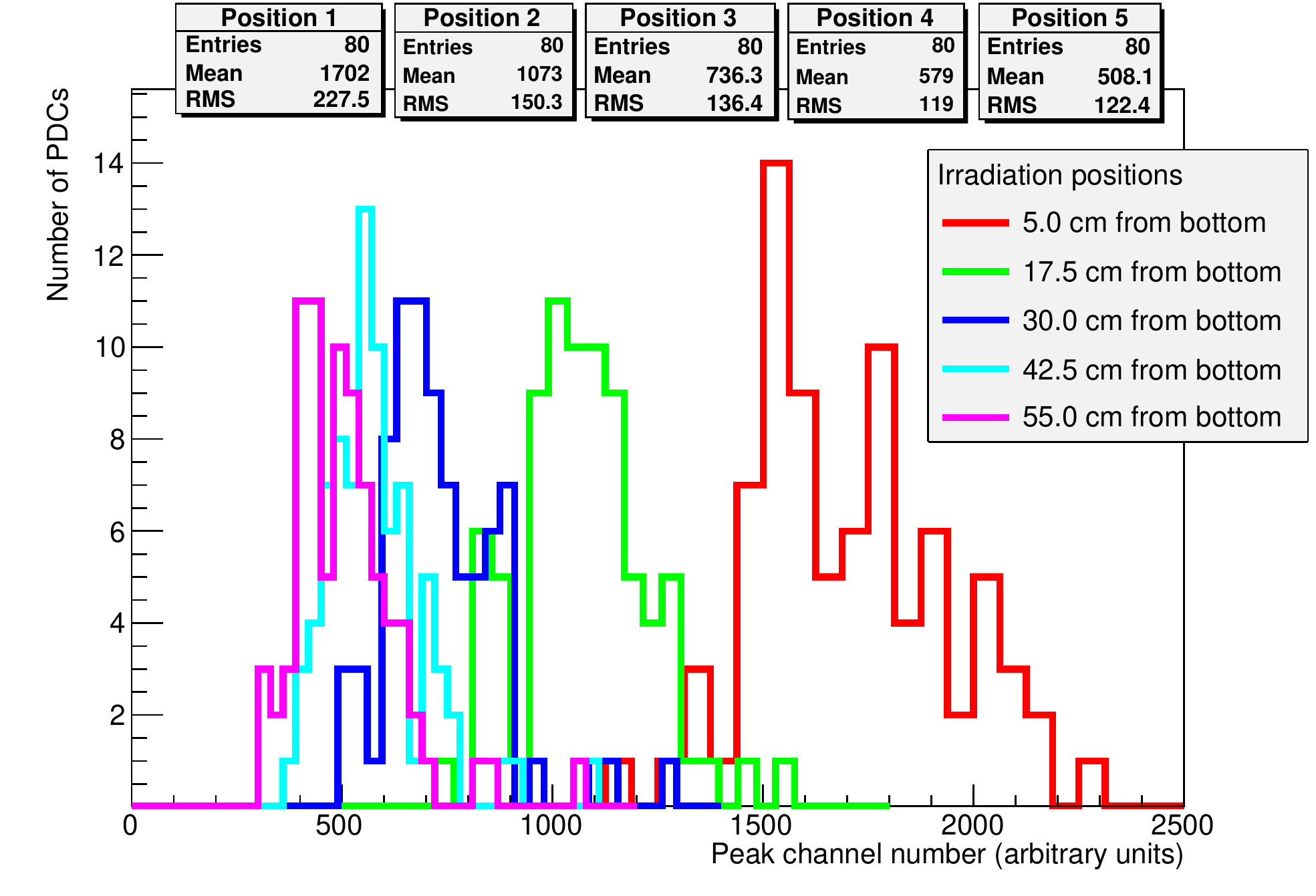}
\caption{\label{lightyield} A summary of light yield measurements for the bottom BGO (top left), the fast scintillator (top right) and for five 
irradiation points on the slow scintillator (bottom).}
\end{figure}

\begin{figure}[!hbt]
\centering
\includegraphics[width=.85\textwidth]{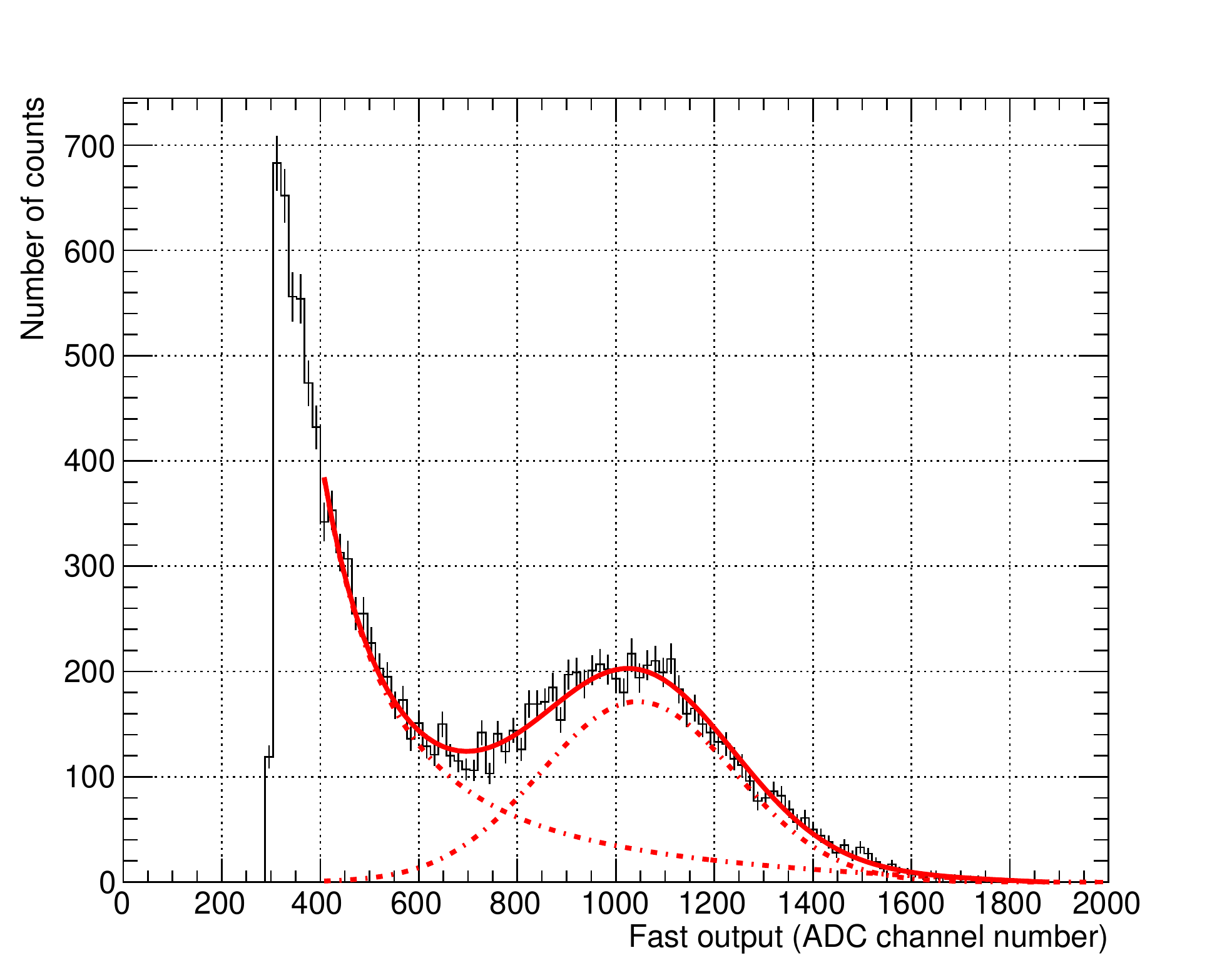}
\caption{\label{PDC calibration example}Example of a PDC calibration spectrum. A clear peak resulting from the absorption of \mbox{59.5 
keV} photons by the fast scintillator can be seen around channel number 1000. The step around channel number 300 corresponds to the 
trigger threshold. The spectrum has been fitted with dashed curves describing the Gaussian signal and exponential background. The thick 
solid line corresponds to the sum of these two curves.}
\end{figure}

A procedure was developed to characterise PDC performance and determine which units would be used for the
flight polarimeter~\cite{Kiss - PhD thesis}:  
(a) the fast plastic scintillator was irradiated from the side using $^{241}$Am (59.5~keV $\gamma$); (b) the slow plastic scintillator was 
irradiated from the side at 5 equi-separated points using $^{90}$Sr (electrons); (c) the BGO crystal was irradiated from the side using 
$^{137}$Cs (662~keV $\gamma$). The distribution of measured light yields are summarised in Figure~\ref{lightyield}. The typical difference 
in light yield at the two ends of the 60~cm long slow scintillator is approximately 75\%. Previous tests showed that the variation in light 
yield along the length of the fast scintillator was 13\%~\cite{Mizuno}.
The position of PDCs within the polarimeter was governed by the light yield results. PDCs with the highest fast scintillator light yield were 
placed at the centre of the scintillator array, while units with high slow scintillator light yield were selected for the outer-most detector ring, 
where background is expected to be more severe. Assembled SAS units were characterised using photons from $^{137}$Cs incident on the 
centre point of each 20~cm long BGO crystal. The light yield was found to vary by approximately \mbox{10\%} over the 60~cm length of a 
SAS element. 
Photomultiplier tube operating voltages were chosen to produce a uniform response to \mbox{59.5 keV} photons photoabsorbed in the fast scintillator array.
An example of a reconstructed energy spectrum is shown in Figure~\ref{PDC calibration example}.
The plastic scintillator response to other photon energies has also been studied using synchrotron photons of various energies in the range between \mbox{6 keV} and \mbox{70 keV}~\cite{Mizuno}. 
Based on this data, a correction can be applied for the non-linearity of the signal amplitude as a function of energy, e.g. at 25 keV, approximately 95\% of the deposited energy is detected.

The polarimeter and data acquisition electronics are housed inside an aluminium pressure vessel with \mbox{3 mm} wall thickness. 
All vessels are purged with dry nitrogen to prevent formation of ice at low temperatures during the balloon flight. The pressurised 
environment (1 atm at room temperature) simplifies the thermal management and selection procedure for components.
X-rays enter the polarimeter through a thin window in the pressure vessel which comprises three materials: 
(i) a \mbox{190~$\mu$m} thick PEEK (polyetheretherketone) window which provides a pressure seal; 
(b) a \mbox{50~$\mu$m} thick DuPont Tedlar light-tight window; and, 
(c) a \mbox{200~$\mu$m} thick aluminised Mylar window to reduce solar heating.
The air column inside the \mbox{60 cm} long slow scintillator tubes constitutes an equivalent atmospheric overburden of less than 
\mbox{0.1 g/cm$^2$}. The transmission through the window materials in 
combination with the air column exceeds \mbox{95\%} for 25~keV photons. 

\begin{figure}[!hbt]
\centering
\includegraphics[width=\textwidth]{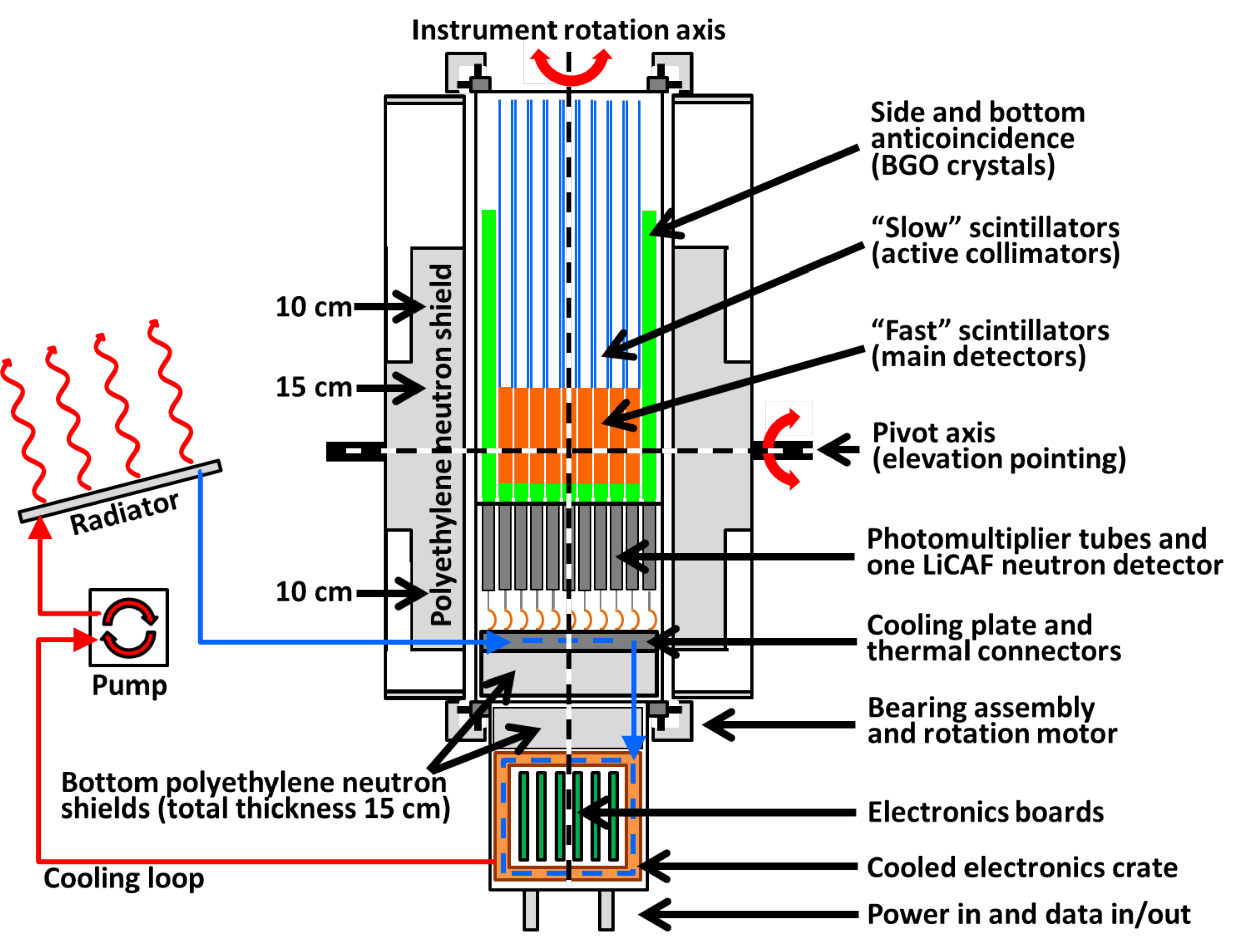}
\caption{\label{PoGOLite instrument sketch}Cross-section of the PoGOLite polarimeter, including the cooling loop. The polarimeter system 
(scintillators, photomultiplier tubes and electronics) is housed inside a system of pressure vessels.}
\end{figure}

In order to meet the desired lower energy threshold for polarimetric measurements, the temperature of the photomultipliers must be actively controlled 
during flight in order to limit photomultiplier dark current.
A fluid-based cooling system~\cite{pau} is used to dissipate the heat generated inside the polarimeter ($\sim$110~W). The heat originates 
from the DC/DC converters inside the tightly-packed array of 91 photomultiplier tubes, with approximately \mbox{$\sim$0.35 W} expended per PMT.
The remaining heat load originiates from the data acquisition electronics. 
Paratherm LR heat-transfer fluid is pumped through a cooling plate inside the polarimeter, which has a thermal path to each PMT. The cooling fluid is also 
circulated through the walls of the housing for the data acquisition electronics. Heat is dissipated as the fluid is circulated through channels in 
a radiator plate which is directed towards the cold sky during the flight. An overview of the polarimeter components and cooling loop is 
shown in Figure~\ref{PoGOLite instrument sketch}.

Atmospheric neutrons are the dominant background to PoGOLite measurements~\cite{MKo thesis}. 
Neutrons can fake valid Compton scattering events through elastic scattering off protons in the fast scintillators. 
This background is reduced by approximately an order of magnitude using a polyethylene shield surrounding the polarimeter assembly. 
In order to reduce the mass of polyethylene needed, the shield has a varying thickness. Its geometry was optimised using Geant4 
simulations where the neutron-induced background as a function of the shield thickness and the signal-induced rate as a function of the 
float altitude (which is mass-dependent) were considered. The implemented shielding configuration, shown in Figure~\ref{PoGOLite instrument sketch}, 
has a maximum thickness of 15~cm and a total mass of $\sim$300 kg. The neutron flux in the vicinity of the fast scintillators
is monitored using a custom LiCAF-based (\mbox{LiCaAlF$_{6}$}) neutron detector~\cite{LiCAF}. 
The detector was used in an independent balloon mission to allow first studies of the high 
latitude stratospheric neutron environment~\cite{pogolino}. 
%%%%%%%%%%%%%%%%%%%%%%%%%%%%%%%%%%%%%%%%%%%%%%%%%%%%%%%%%%%%%%%%%%%
%
\section{Polarimeter data acquisition system}
\label{Polarimeter electronics and control systems}
\begin{figure}[!hbt]
\centering
\includegraphics[width=\textwidth]{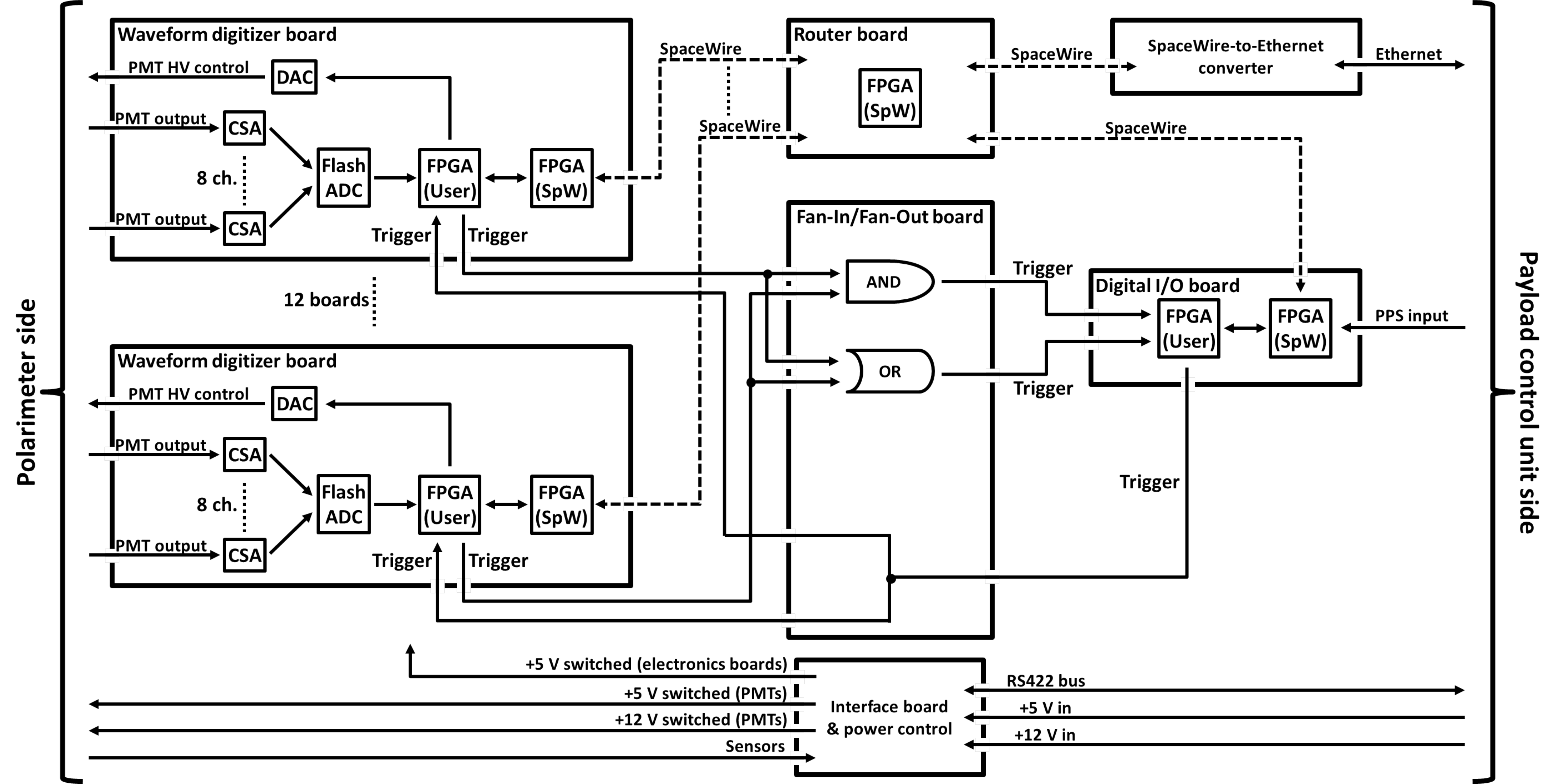}
\caption{\label{DAQ overview}Overview of the PoGOLite data acquisition electronics. For simplicity, only two of the twelve waveform digitizer boards 
are shown. Dashed lines indicate SpaceWire~\cite{SpaceWire} interfaces, while solid lines correspond to LVDS connections, unless 
otherwise indicated. All SpaceWire boards are provided by Shimafuji Electric. An interface and power control board is used to control the 
power to the data acquisition boards and photomultiplier tubes and collects signals from temperature and pressure sensors. The reference 
voltage for each PMT, which governs the internal DC/DC conversion for the dynode high-voltage steps, can be individually regulated using 
digital-to-analog converters (DACs) on each waveform digitizer board.}
\end{figure}

\begin{figure}[!hbt]
\centering
\includegraphics[width=.85\textwidth]{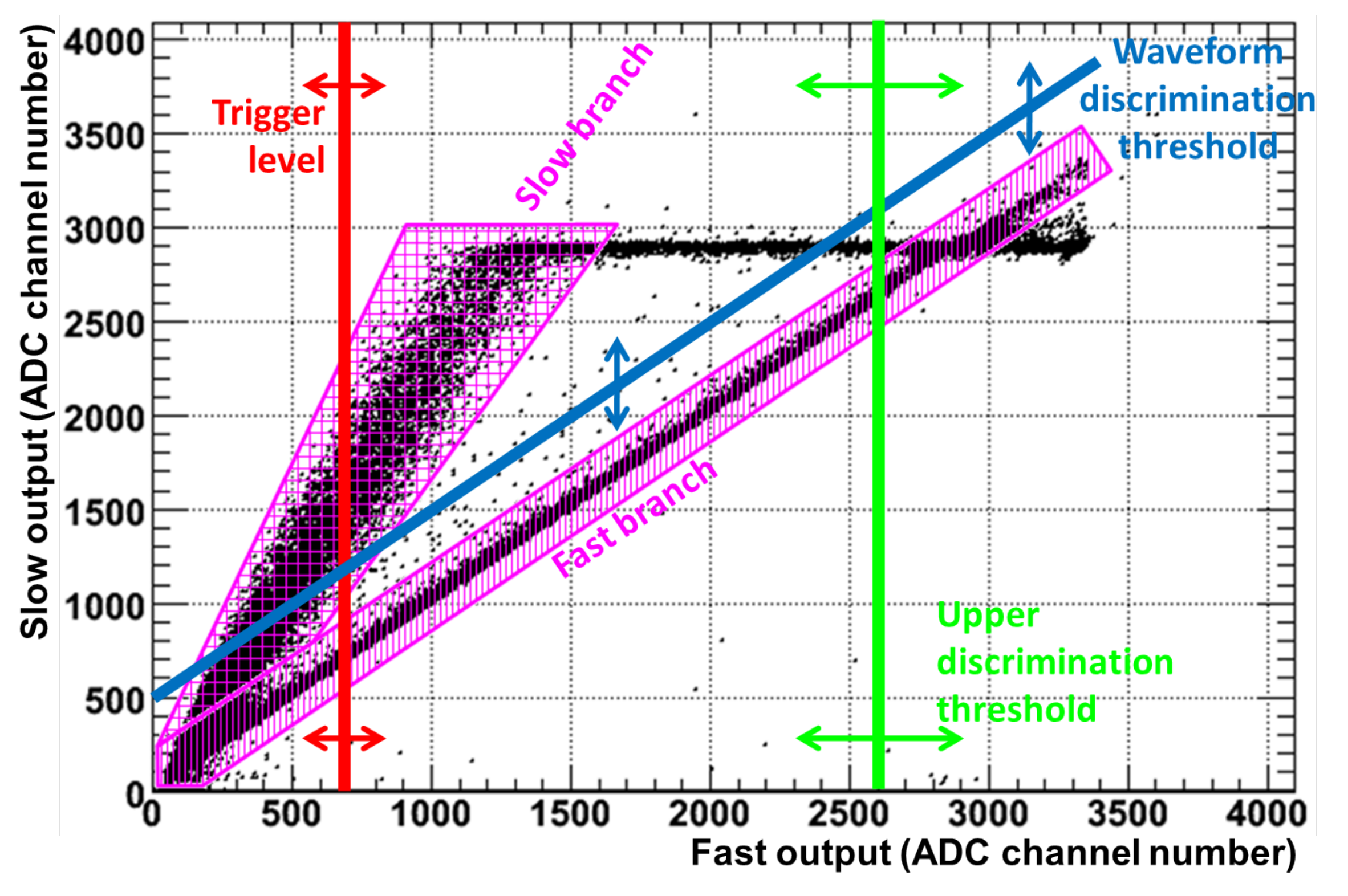}
\caption{\label{Threshold description}An illustration of the relationship between fast and slow outputs from one PDC. Each point corresponds to one recorded waveform. 
Data points are separated into two branches, corresponding to events from the fast scintillator and slow scintillator/BGO, respectively. 
Photons scattering from one component to the other yield a mixed 
component waveform that results in a point between the two branches. The flattening around channel 2800 on the vertical axis is caused by 
saturation. The fast output is a measure of the deposited energy, and 1000~channels corresponds to approximately \mbox{60 keV} (see Figure~\ref{PDC calibration example}). 
The function of the trigger and threshold settings is illustrated.}
\end{figure}

A schematic overview of the data acquisition electronics is shown in Figure~\ref{DAQ overview}. 
Polarimeter photomultiplier tube signals are fed into charge-sensitive amplifiers on the waveform digitizer boards, the outputs of which are continuously 
sampled by flash analog-to-digital converters with \mbox{12 bit} precision at \mbox{37.5 MHz}. For each sample, the signal amplitude is calculated as the difference between the 
current sample point value and the value three samples earlier.
A trigger is issued if the signal amplitude exceeds a threshold level, resulting in all waveforms with an amplitude exceeding a lower hit threshold being 
recorded with 50 sample points during \mbox{$\sim$1.3 $\mu$s}. The stored waveform starts approximately {0.4 $\mu$s} before 
the trigger. This allows offsets in the waveform baseline, e.g. due to large preceding signals or PMT after-pulses, to be identified when determining
the signal amplitude.
\begin{figure}[!hbt]
\centering
\includegraphics[width=.48\textwidth]{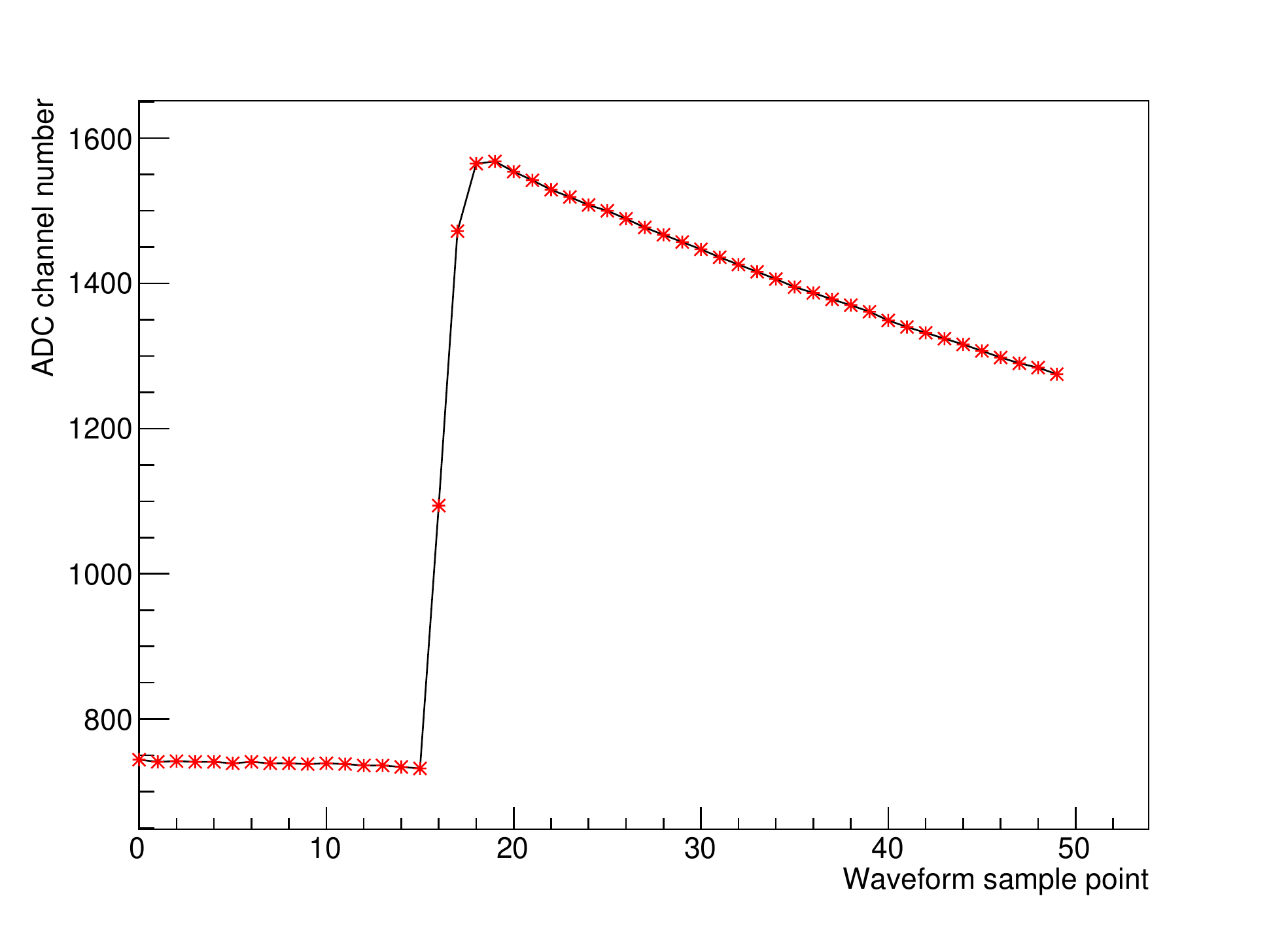}  %.495\textwidth for same line
\includegraphics[width=.48\textwidth]{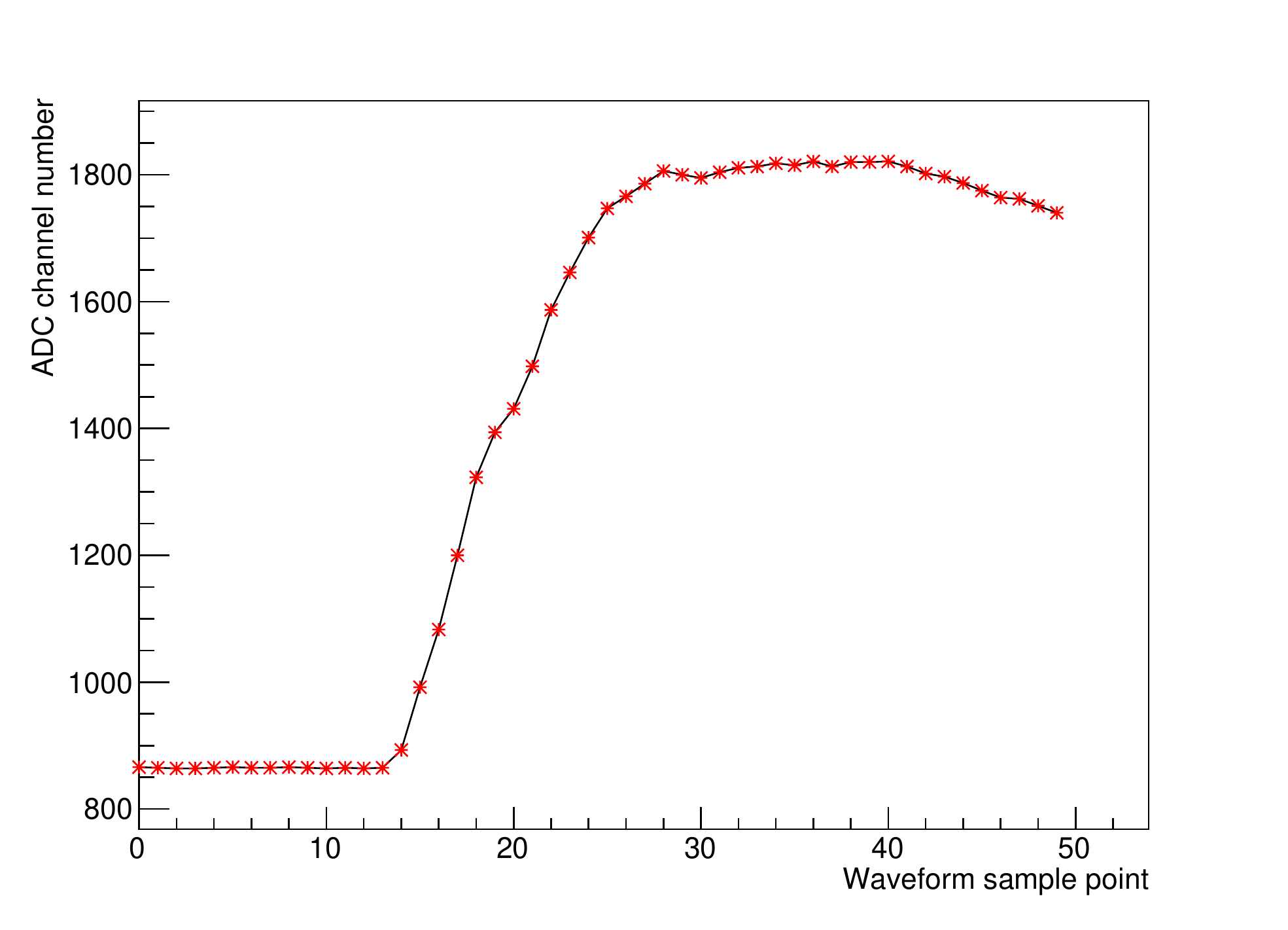}  %.495\textwidth for same line 
\caption{\label{Fast slow pulse examples}Waveform from the fast scintillator (left) and from the slow scintillator or BGO crystal (right). The 
nominal baseline is around ADC channel number~800, and can fluctuate both high (riding on previous event) or low (undershoot during 
recovery from a preceding pulse). The pre-trigger sample points allow such offsets to be corrected.}
\end{figure}
Typical waveforms from the fast scintillator and slow/BGO scintillator are shown in 
Figure~\ref{Fast slow pulse examples}. 
Pulse shape discrimination uses a calculated fast (slow) output amplitude defined as the maximum pulse height considered within a sliding 
integration window 3 (14) samples wide.
Events from the fast scintillator and slow/BGO scintillator can be separated into two branches,
as shown in Figure~\ref{Threshold description}.

Each of the PMT input channels is subject to threshold settings, as illustrated in Figure~\ref{Threshold description}.
The {\it trigger threshold} defines the minimum PDC signal amplitude required to initiate data acquisition. The threshold value is set to \mbox{300 ADC} channels (\mbox{15 - 20 keV}, 
depending on the PDC), i.e. compatible with a photoelectric absorption. 
Once a valid trigger has been received, all PDCs with a fast or slow output amplitude exceeding the {\it hit threshold} will 
be read out. This threshold level is conservatively set to \mbox{$\sim$0.5 keV} in order to accept the majority of low energy Compton events, while 
still rejecting electronics noise.
The {\it waveform discrimination threshold} is a simple pulse shape discrimination selection, allowing slow
scintillator or BGO hits to be rejected. 
An {\it upper discrimination threshold} (UD) is used to reject large amplitude signals, e.g. from minimum ionising cosmic rays 
interacting within a scintillator. In this case, data acquisition dead time ($\sim$11 $\mu$s) is introduced to 
allow the waveform baseline to recover before further triggers are accepted. 
Algorithms implemented in field-programmable gate arrays on the waveform digitizer boards search for signal amplitudes compatible with a trigger. 
If coincident waveform/upper discrimination signals 
from other channels are identified during a 16~clock cycle window (\mbox{$\sim$0.4 $\mu$s}) the event is vetoed, otherwise a data acquisition request is
issued to all waveform digitizer boards and waveforms from all channels with a signal amplitude above the hit threshold are stored. 
Requiring that upper discrimination and waveform discrimination threshold conditions are fulfilled significantly reduces polarimeter dead-time and 
data volume. To ensure that the thresholds do not introduce a bias in the measurements, data can also be recorded without 
applying thresholds. This has been tested for various source and background configurations~\cite{Ground calibration paper}.
As well as storing the sampled waveforms, minute-wise signal amplitude histograms are recorded for each input channel with a \mbox{10 bit} accuracy.
The histograms are filled independently of the trigger and threshold settings for signal amplitudes above a {\it histogram threshold} ($\sim$0.5~keV).
These files are used to calculate e.g. the rate of saturated (UD)
events or the rate of background events.
After vetoes, the amount of compressed scientific data produced during a one hour Crab observation is $\sim$350~MBytes. 

%
%%%%%%%%%%%%%%%%%%%%%%%%%%%%%%%%%%%%%%%%%%%%%%%%%%%%%%%%%%%%%%%%%%%
%
\section{Control and data storage systems}
\label{Control electronics}
\begin{figure}[!hbt]
\centering
\includegraphics[width=\textwidth]{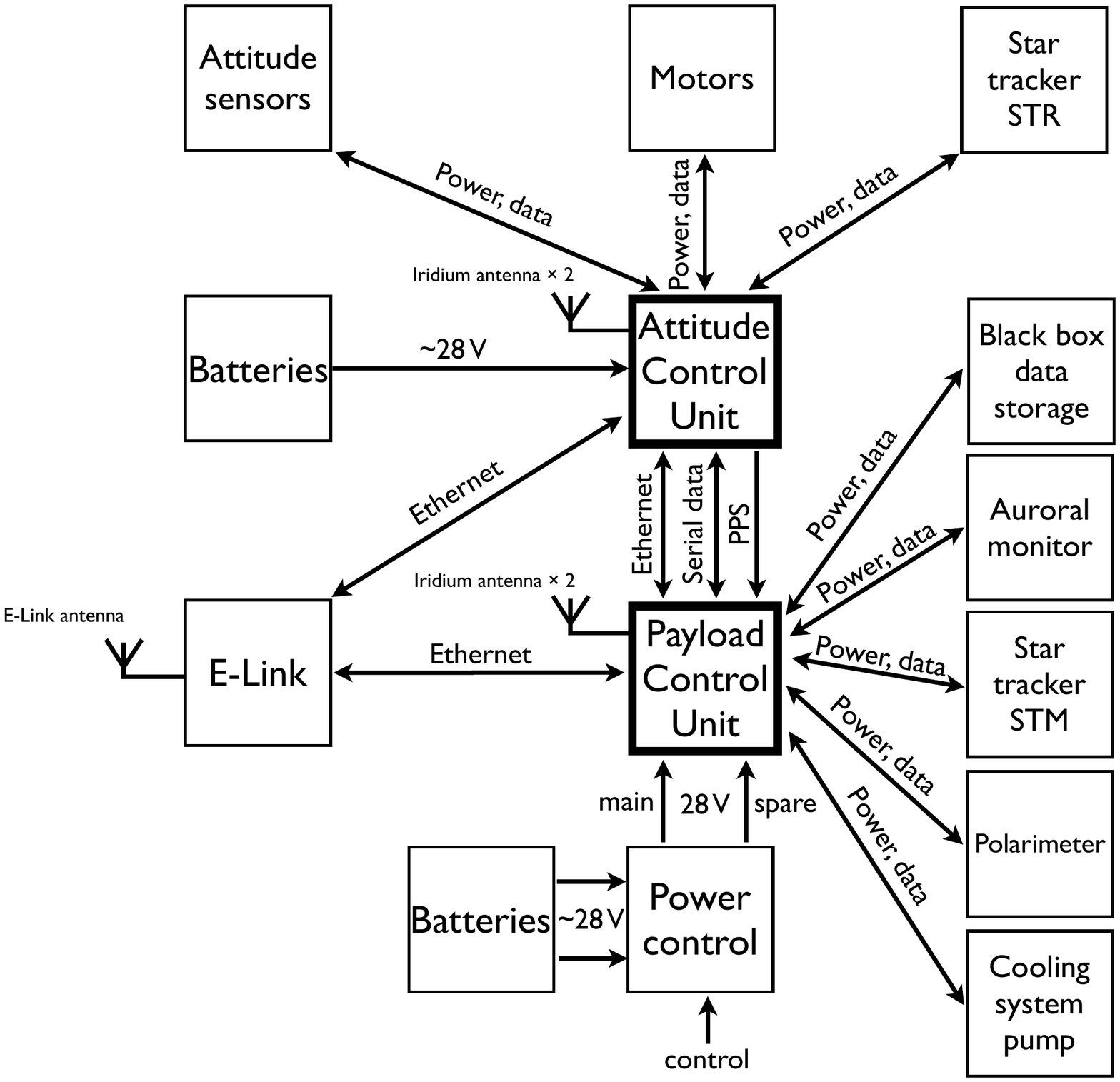}
\caption{\label{Electronics overview} Overview of the PoGOLite systems and their interconnections. The Payload Control Unit (PCU) 
commands the polarimeter, cooling system pump, STM star tracker and data storage unit. The auroral monitor is an independent piggyback 
instrument. The Attitude Control Unit (ACU) interfaces to a number of motors and attitude sensors. The PCU and ACU communicate via a 
serial bus and Ethernet. Both systems are connected to the E-Link and Iridium systems for communication with the ground.}
\end{figure}

Two PC/104-standard flight control computers within the Payload Control Unit (PCU) provide redundant control of the polarimeter
and other systems, as illustrated in Figure~\ref{Electronics overview}.
A dedicated custom real-time computer provides an interface to data from the Attitude Control Unit (ACU) and 
controls power distribution to the polarimeter. A pulse per second (PPS) signal from a GPS unit in the ACU is read by the polarimeter data acquisition system. 
Receipt of this signal generates timestamped waveform data across all waveform digitizer boards which can be synchronised to absolute GPS time. The relative 
offset from the GPS second transition for each polarimeter event is determined using a \mbox{37.5 MHz} crystal oscillator on each waveform digitizer 
board. The absolute polarimeter timing precision is \mbox{$\sim$1 $\mu$s}~\cite{Ground calibration paper}. 
Communication with the flight control computers uses the "Ethernet-over-radio", E-Link, system~\cite{E-LINK} which operates while there 
is line-of-sight to Esrange or a relay station in And{\o}ya, Norway. The link bandwidth ($\sim$1 Mbit/s) allows all scientific data to be downloaded. 
The Iridium satellite communication system is used for over-the-horizon communications. 
The PoGOLite Pathfinder is designed to operate autonomously during the Iridium phase, determining the best suitable observation target 
based on parameters such as time, position, altitude and target priority~\cite{Miranda}.  
Since the Iridium bandwidth is extremely limited, \mbox{$\sim$1 kbit/s}, only
a limited subset of processed scientific data can be downloaded. Post-flight recovery of the data storage units is therefore required for a 
full waveform-based analysis.
Data is stored on solid state disk arrays in the PCU and in an independent data storage unit which has an inbuilt computer and provides a redundant back-up for the flight control computers. 
Scientific and housekeeping data are automatically synchronised between these three computers during the flight.

%
%%%%%%%%%%%%%%%%%%%%%%%%%%%%%%%%%%%%%%%%%%%%%%%%%%%%%%%%%%%%%%%%%%%
\section{Attitude control system}
\label{ACS}
A custom attitude control system 
keeps the viewing axis of the polarimeter aligned to the sidereal motion of observation 
targets and compensates for perturbations such as torsional forces in the flight-train rigging and stratospheric winds. In order to prevent 
performance degradation due to shadowing effects of the slow scintillator collimators onto the fast scintillator array, pointing to within 
\mbox{$\sim$0.1$^{\circ}$} is required during observations~\cite{Cecilia}. 

The polarimeter assembly is mounted in a gimbal. 
A number of digital direct drive torque motors~\cite{JES_ESAPAC} are mounted on the gimbal and used to orientate the polarimeter.
One motor acts directly on the polarimeter elevation axle.  
Azimuthal pointing is achieved using a combination of a motor-driven flywheel system and a motor which acts upon the coupling to the 
flight-train rigging (which connects the gimbal to the balloon).
The flywheel provides high frequency compensation, while the balloon train motor provides low frequency and static torque compensation.
If the flywheel speed saturates, angular momentum can be transferred to the balloon 
using the flight-train motor.
The use of direct drive torque motors eliminates problems with gear-box inertia and backlash. 
The polarimeter roll motor is controlled from the attitude control system. A standard geared motor is used, to match the moderate 
performance requirements and manage the high and fluctuating torque required to turn the polarimeter.

Digital control signals are passed between the motors and a real-time processor which receives inputs from a number of attitude 
sensors mounted on the gimbal.
The absolute gondola heading is primarily determined by a differential GPS (DGPS) unit. 
Two antennas are separated by $\sim$10~m and the phase difference between the GPS carrier wave signals is used to determine a relative 
positioning of the two antennas to within $\sim$10~mm accuracy. 
Heading values are generated at a relatively slow rate of 20~Hz but with high accuracy, $\sim$0.05$^\circ$. A three-axis magnetometer is 
used as a back-up heading sensor, with an accuracy of a few degrees. 
The elevation of the polarimeter is gauged with an accuracy better than 0.01$^\circ$ with respect to the gimbal structure by encoders 
mounted on the elevation axle. 
An inclinometer is mounted on the gimbal to measure the absolute attitude of the gondola with respect to the gravity of the Earth. 
An Intertial Measurement Unit (IMU) is used to correct for high frequency disturbances coming from the balloon train, friction and cabling. 
The IMU comprises three orthogonally oriented micromechanical gyro sensors and accelerometers mounted on the polarimeter elevation axis. 
Algorithms implemented in the IMU firmware use a combination of inputs from all
other attitude sensors to compute a wide bandwidth high accuracy attitude estimate.
 
\begin{figure}[!hbt]
\centering
\includegraphics[width=\textwidth]{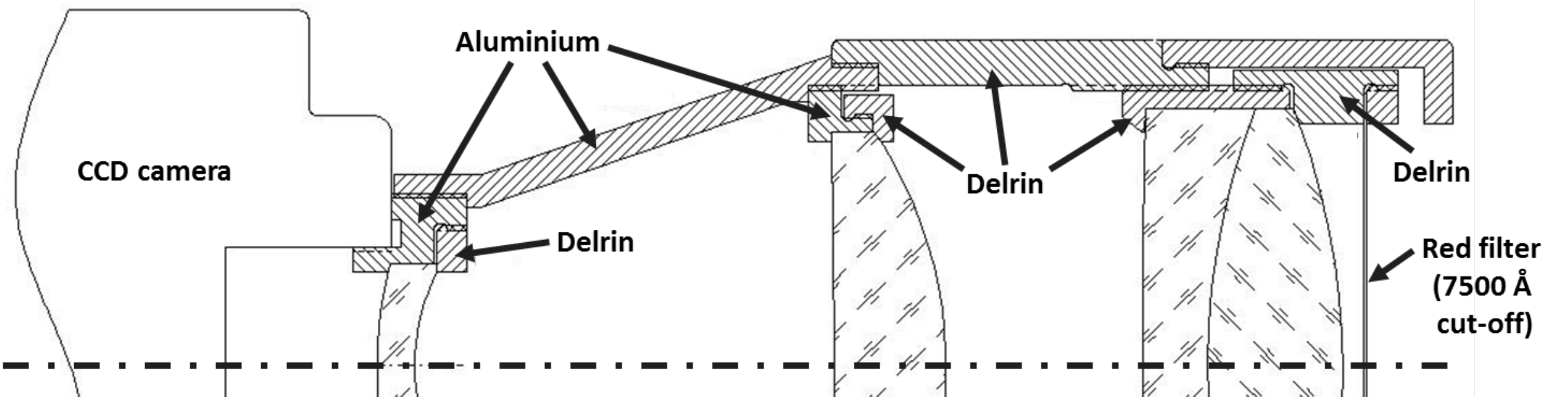}
\caption{\label{STR mechanical design}Mechanical design of the STR star tracker camera. The choice of materials circumvents the need for 
active thermal management or refocusing optics.}
\end{figure}

The polarimeter field-of-view is positioned on the star-field of interest using signals from the DGPS and IMU to close the attitude control 
loop. 
The star fields are imaged by two star trackers, ``STM'', narrow field-of-view, \mbox{$2.6^{\circ} \times 1.9^{\circ}$} 
and ``STR'', wide field-of-view, \mbox{$5.0^{\circ} \times 3.7^{\circ}$}.  
The STM star tracker~\cite{Cecilia} is based on a flight-proven design~\cite{BLAST} \cite{Dietz}.
The STR star tracker is significantly more compact and light-weight than the STM, and is foreseen to be used for future balloon 
campaigns. 
STR uses the same \mbox{1392 $\times$ 1024}~pixel CCD camera as STM, but has a custom 
temperature-compensated fixed-focus design~\cite{LODAC}. The design concept is illustrated in Figure~\ref{STR mechanical design}. To 
protect the CCD cameras during pointing operations, both star trackers are equipped with mechanical shutters triggered automatically by 
photo-diodes if the polarimeter is inadvertently pointed at the Sun.  

The desired pixel position of a guide star is calculated based on the gondola location, time and observation target coordinates. 
Using pixel values from a small part of the CCD image the measured guide star position is used to close the attitude control loop. 
Stability constraints, hot-pixel maps and neighbouring pixel values are used to filter erroneous pixel readings, caused, for example, by cosmic 
rays. 
For Crab observations, the guide star is HIP~26451 (visual magnitude M=2.95) for STR and HIP~26328 (M=7.00) for STM. 

%%%%%%%%%%%%%%%%%%%%%%%%%%%%%%%%%%%%%%%%%%%%%%%%%%%%%%%%%%%%%%%%%%%
\section{Gondola}
\label{gondola}
\begin{figure}[!hbt]
\centering
\includegraphics[width=\textwidth]{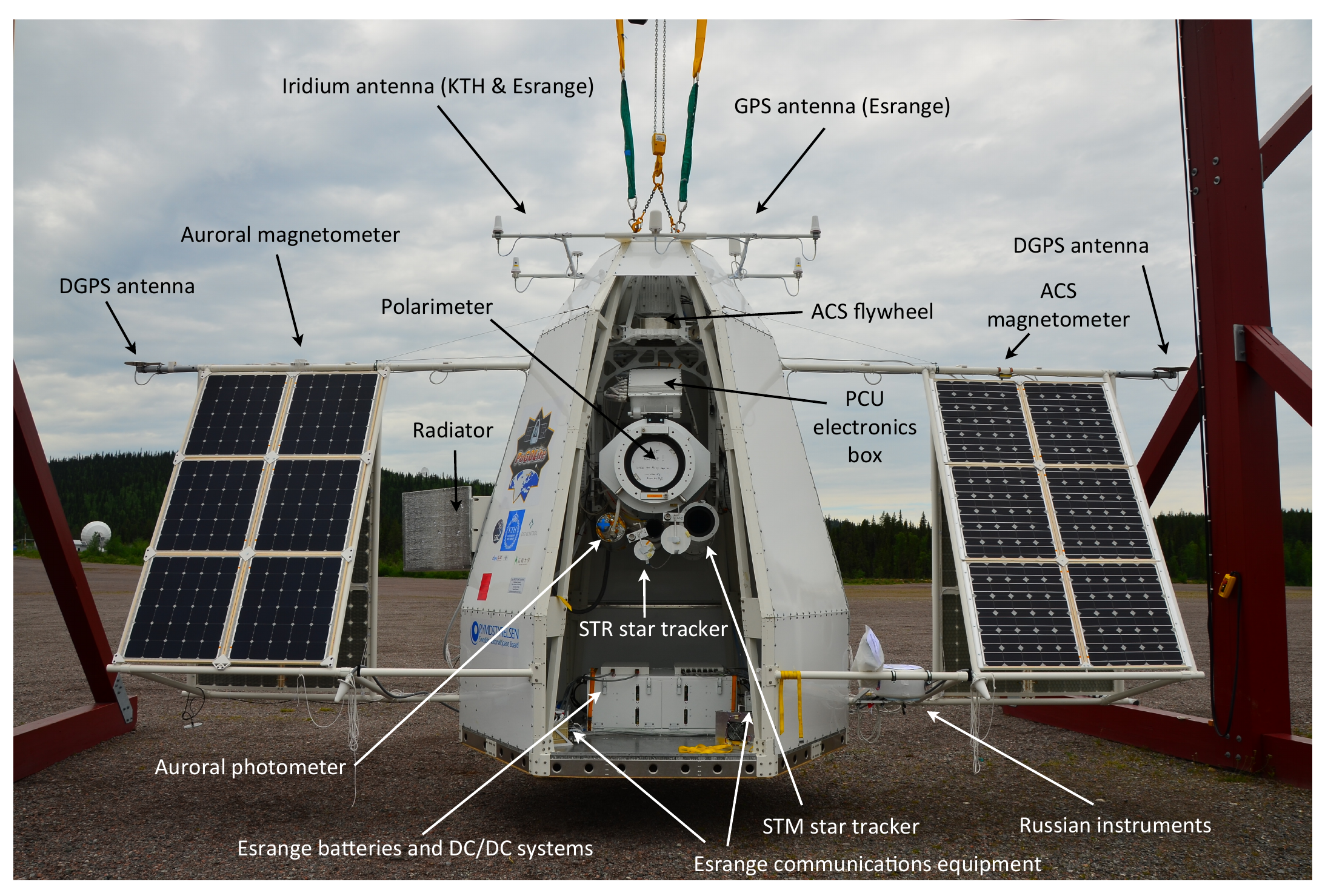}
\caption{\label{Gondola}The PoGOLite gondola (5 m in height) during ground-based tests at Esrange. Before flight, crash pads and ballast 
hoppers are mounted under the gondola. The opening at the front of gondola is closed up to the level of the star trackers during flight to 
prevent solar radiation from reaching components inside the gondola.}
\end{figure}

The gimbal is mounted inside a gondola structure developed by SSC Esrange Space Centre, as shown in Figure~\ref{Gondola}.
The gondola reduces the solar flux incident on the polarimeter and ancillary equipment and protects the polarimeter from damage during 
landing. The gondola comprises a tubular aluminium structure covered in composite honeycomb panels which affords a light-weight but rigid 
structure able to withstand the \mbox{10 G} shocks expected during parachute deployment and landing. 
Solar panels are located on each side of the gondola (forward and backward-facing), giving a total area of \mbox{12 m$^2$} on each side, 
providing approximately \mbox{1 kW} charging power at float altitude. The overall width of the gondola is \mbox{10 m}, for a total height 
of \mbox{5 m}. The suspended weight is approximately \mbox{1850 kg} (excluding ballast), which breaks down as (i) polarimeter, 600~kg; (ii) 
attitude control system including gimbal, 300~kg; (iii) and gondola, including power supply, solar panels, communications and ancillary 
equipment, 950~kg. The gondola is suspended approximately 100~m under a helium-filled 
polyethylene balloon with a volume of 1.1~Mm$^3$, corresponding to a fully inflated diameter of approximately 150~m. Neither the gondola or balloon 
occludes the polarimeter field-of-view during observations up to an elevation of 60$^{\circ}$.
%
%%%%%%%%%%%%%%%%%%%%%%%%%%%%%%%%%%%%%%%%%%%%%%%%%%%%%%%%%%%%%%%%%%%
%
\section{The 2013 near-circumpolar flight}
\label{The 2013 flight}
%
%%%%%%%%%%%%%%%%%%%%%%%%%%%%%%%%%%%%%%%%%%%%%%%%%%%%%%%%%%%%%%%%%%%
\subsection{Flight overview}
\begin{figure}[!hbt]
\centering
\includegraphics[width=\textwidth]{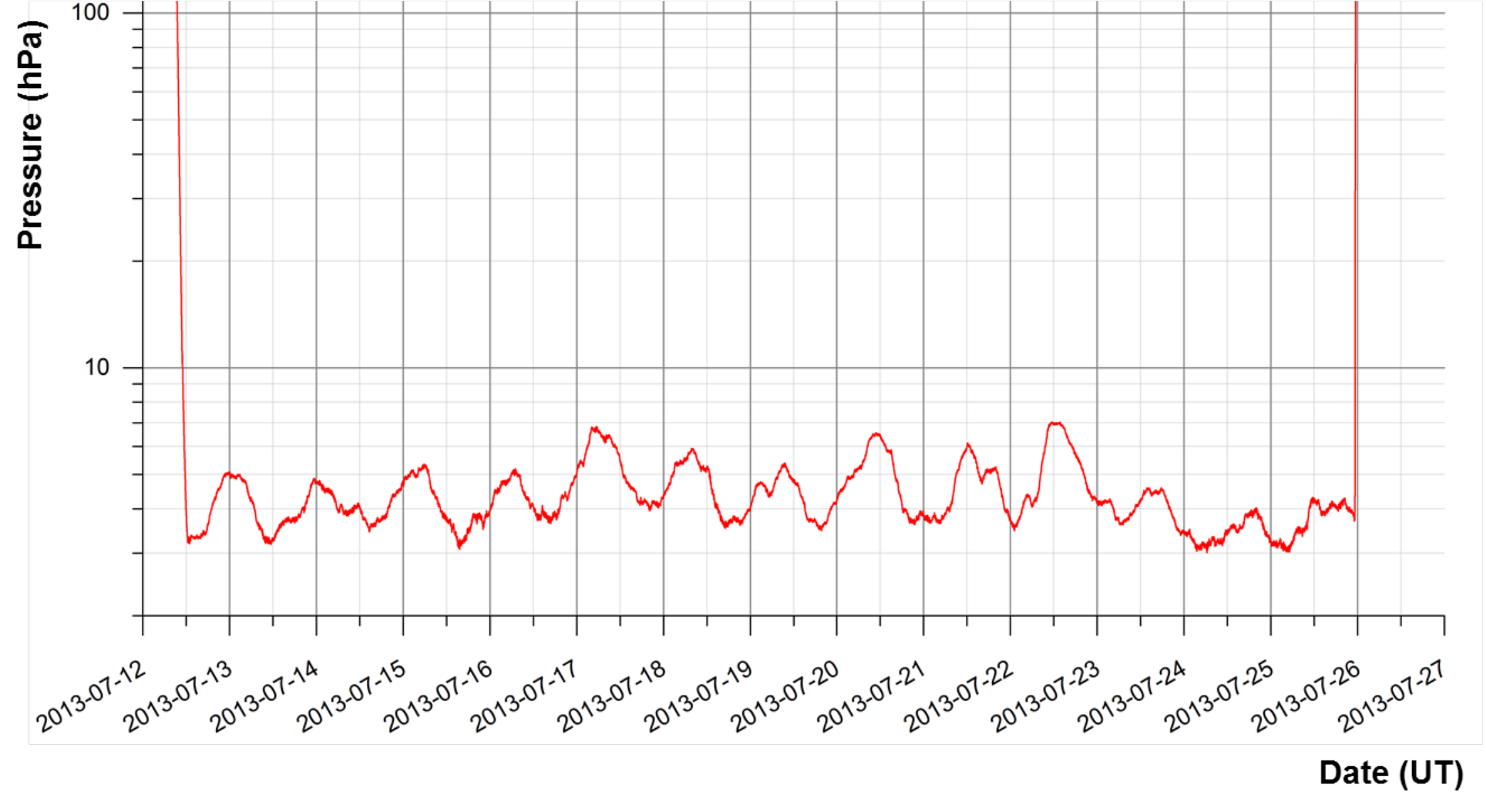}
\caption{\label{Flight altitude}Ambient pressure in hectopascal (\mbox{1 hPa $\approx$ 1.02 g/cm$^2$}) throughout the flight. Diurnal 
variations are seen, caused by heating and cooling of the helium inside the balloon. The initial float altitude is \mbox{39.9 km}. Towards the 
end of the flight, when some ballast has been released, the altitude exceeds \mbox{40.0 km}. Data courtesy of SSC Esrange Space Centre.}
\end{figure}

\begin{figure}[!hbt]
\centering
\includegraphics[width=0.7\textwidth]{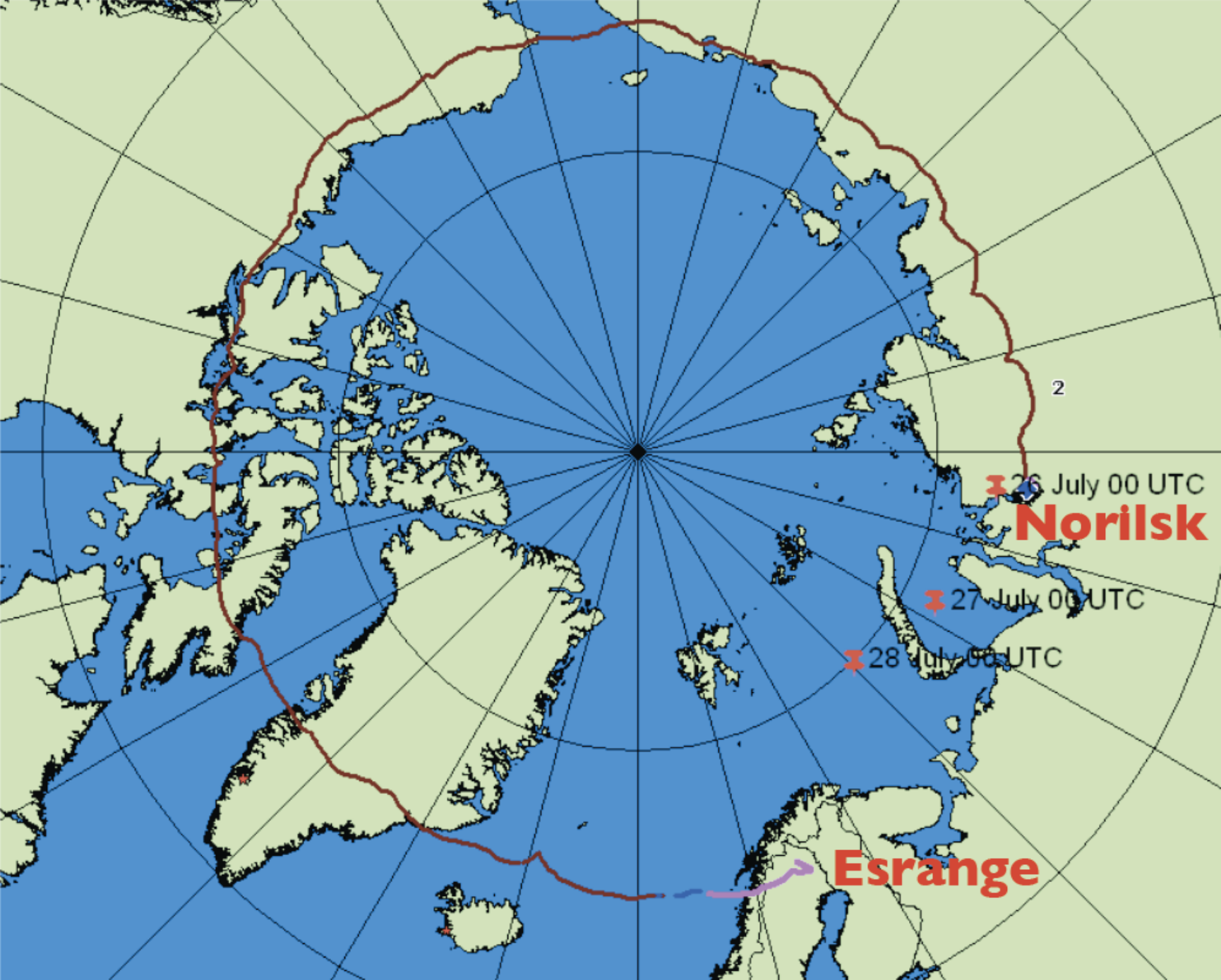}
\caption{\label{Flight trajectory}The near-circumpolar flight trajectory, starting from Esrange in northern Sweden and moving in a westerly 
(clockwise) direction. After launch, E-Link coverage lasted for 19 hours as shown in purple (Esrange station) and blue (And{\o}ya station). 
The periodic pattern in the trajectory, seen towards the end of the flight, results from ballast drops aimed at halting 
the tendency for the balloon to move in a northerly direction. 
Predictions for the gondola position on July 26th, 27th and 28th are indicated by red markers. Since the predicted trajectory would miss 
Scandinavia, the flight was terminated near the city of Norilsk. Data courtesy of SSC~Esrange Space Centre.}
\end{figure}

The PoGOLite Pathfinder balloon campaign was conducted at the Esrange Space Centre (\mbox{68.89$^\circ$ N}, \mbox{21.11$^\circ$ E}), 
near the Swedish town of Kiruna. Towards the end of May, stratospheric winds start to flow in a westerly direction, providing 
long duration flight opportunities until the stratospheric winds start to break down at the end of July. The launch window for PoGOLite 
opens on July~1st when the angular separation between the Sun and Crab exceeds \mbox{15$^{\circ}$}. 
On \mbox{July 12th}, at 08:18~UT, the PoGOLite Pathfinder 
was successfully launched on a planned full circumpolar flight. 
The ascent to float altitude, \mbox{$\sim$39.9 km}, took approximately five hours. 
Observations cycled between the Crab and Cygnus X-1, which was not in the required hard state but allowed observation conditions 
for possible future flights to be studied. 
While passing over Russia, the stratospheric winds started to move in a northerly direction. Predictions based on the balloon position, 
airspeed and direction indicated that the balloon would pass north of Scandinavia.  The flight was therefore terminated near Norilsk, 
\mbox{3000 km} East of  Moscow, where infrastructure was available for the gondola recovery. The gondola was cut from the balloon at 
23:24~UT on July 25th. Ambient pressure measured throughout the flight is presented in Figure~\ref{Flight altitude} and
the balloon trajectory is shown in Figure~\ref{Flight trajectory}. 
The polarimeter assembly and data storage units were returned to Stockholm on January~21st 2014. Inspection of the polarimeter indicated 
that no significant damage had incurred during flight termination or transport.
%
%%%%%%%%%%%%%%%%%%%%%%%%%%%%%%%%%%%%%%%%%%%%%%%%%%%%%%%%%%%%%%%%%%%
%
\subsection{Thermal environment}
\label{sec:thermal}
The thermal environment during summer circumpolar stratospheric balloon flights is characterised by solar illumination which is 
continuous but variable. Due to the low ambient pressure the primary heat transfer mechanism is radiation. A full treatment of the thermal 
environment requires solar radiation, the albedo heat flux from the Earth, coupling to the balloon, radiation to space and convection effects 
to be taken into account. 
During ascent, temperatures of \mbox{$-$40 $^{\circ}$C} were experienced during passage through the tropopause at an altitude of $\sim
$15~km.
Once at float altitude, the thermal environment for the polarimeter systems is primarily dictated by the azimuthal pointing of the gondola. 
During observations of the Crab, the gondola opening is directed towards the Sun. Observations of Cygnus X-1 are in the anti-Sun 
direction.  The temperature of a gondola wall was measured to be \mbox{85 $^{\circ}$C} while solar-illuminated and  \mbox{$-$10 $^{\circ}$C} 
while in eclipse~\cite{privESC}.

During Crab observations, the polarimeter and PCU box are exposed to solar radiation. The polarimeter heating time constant is relatively 
long due to the shielding effects of the thick polyethylene shield and the reflective Mylar window placed over the polarimeter aperture. 
The polarimeter fluid cooling system ensured that data acquisition electronics and photomultiplier temperatures remained within 
specification.
During a Crab observation, the temperature of the waveform digitizer boards was found to increase by $\sim$15 $^{\circ}$C, which affected
the performance of the charge sensitive amplifiers as discussed in section~\ref{sec:obs}.  
The photomultiplier temperature increased by \mbox{$\sim$8 $^{\circ}$C}. The 
effect on the photomultiplier gain can be gauged and compensated for by studying the position of the single photoelectron peak. 
BGO scintillator temperatures in the side anticoincidence system were observed to increase by $\sim$10$^{\circ}$C.
The corresponding decrease in light yield of $\sim$12\% has negligible effect on the efficiency of the anticoincidence 
shield considering the threshold settings adopted. The temperature-induced change in the scintillation light decay time also has no effect.
While the close packed array of plastic scintillators was not equipped with temperature sensors, the temperature dependence of plastic 
scintillator light yield is relatively weak -- at \mbox{60$^{\circ}$C}, the light yield is 95\% of that at \mbox{20$^{\circ}$C}. 

The PCU box is painted white to reflect incoming solar radiation, but unlike the polarimeter, it is not actively cooled. 
The thermal design of the PCU proved inadequate since the flight control computers overheated during Crab observations which 
significantly complicated operations. 
This problem could, however, be partly offset by rotating operations between the flight control 
computers and the data storage computer located on the gondola floor (also prone to overheating when solar-illuminated). 
After the third Crab observation, on July 14th, the power to the polarimeter control system was cycled in order to return to a well-defined state after 
overheating issues. This resulted in a failure of the real-time computer system in the PCU and polarimeter operations were no longer possible. 
Post-flight analysis revealed that the computer file system had become corrupted (unrecoverable during flight). Sensor data showed that the failure was not 
temperature related. 
The ACU can be operated independently from the PCU, which allowed pointing operations to continue.

%%%%%%%%%%%%%%%%%%%%%%%%%%%%%%%%%%%%%%%%%%%%%%%%%%%%%%%%%%%%%%%%%%%
\subsection{Attitude control system performance}
\begin{figure}[!hbt]
\centering
\includegraphics[width=\textwidth]{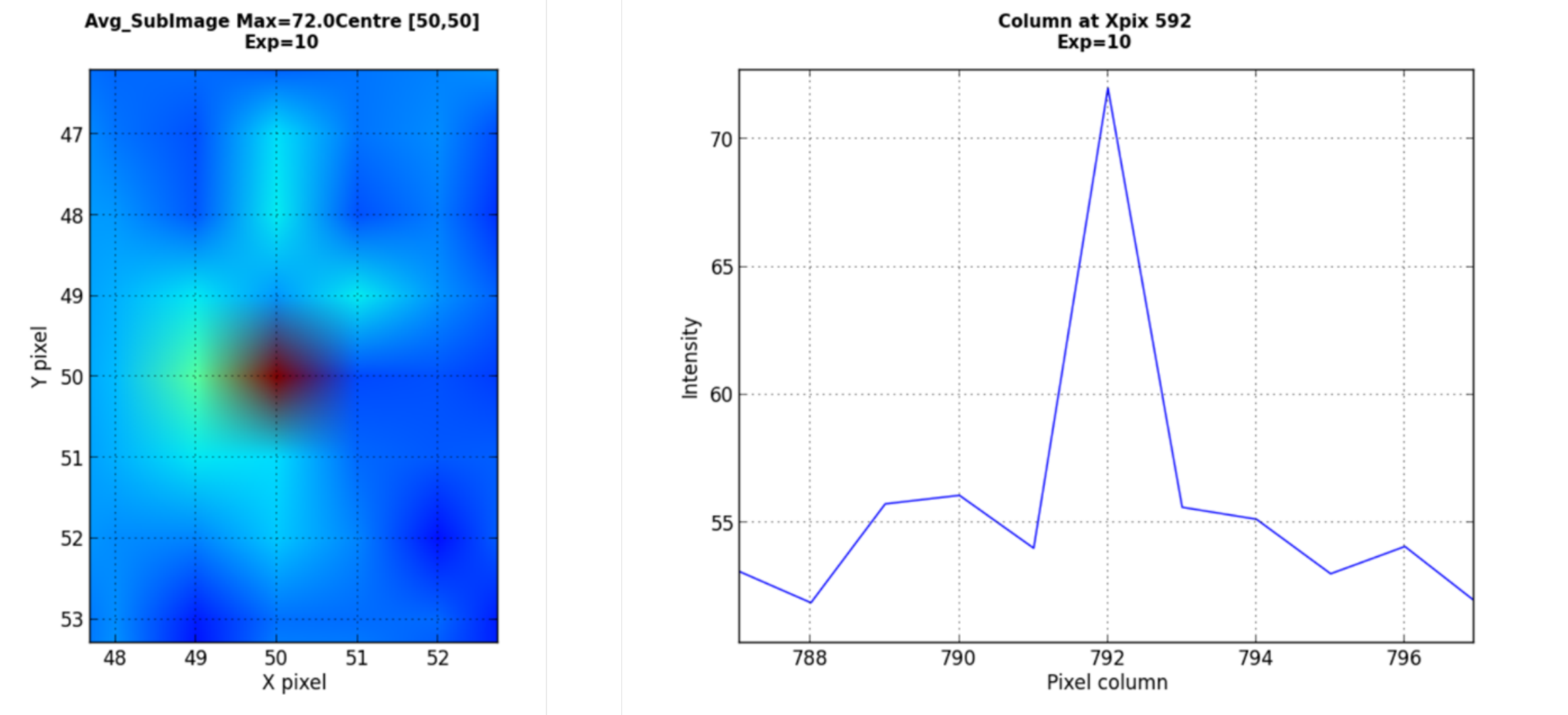}
\caption{\label{Co-added short exposure}Left: sixteen co-added and centered STR images with short exposure (\mbox{10 ms} each). Right: 
averaged vertical pixel values (at pixel column~592) for these images. The symmetric point-spread function indicates that gondola motion and not focus is likely the 
degrading factor for the optical quality.}
\end{figure}

\begin{figure}[!hbt]
\centering
\includegraphics[width=\textwidth]{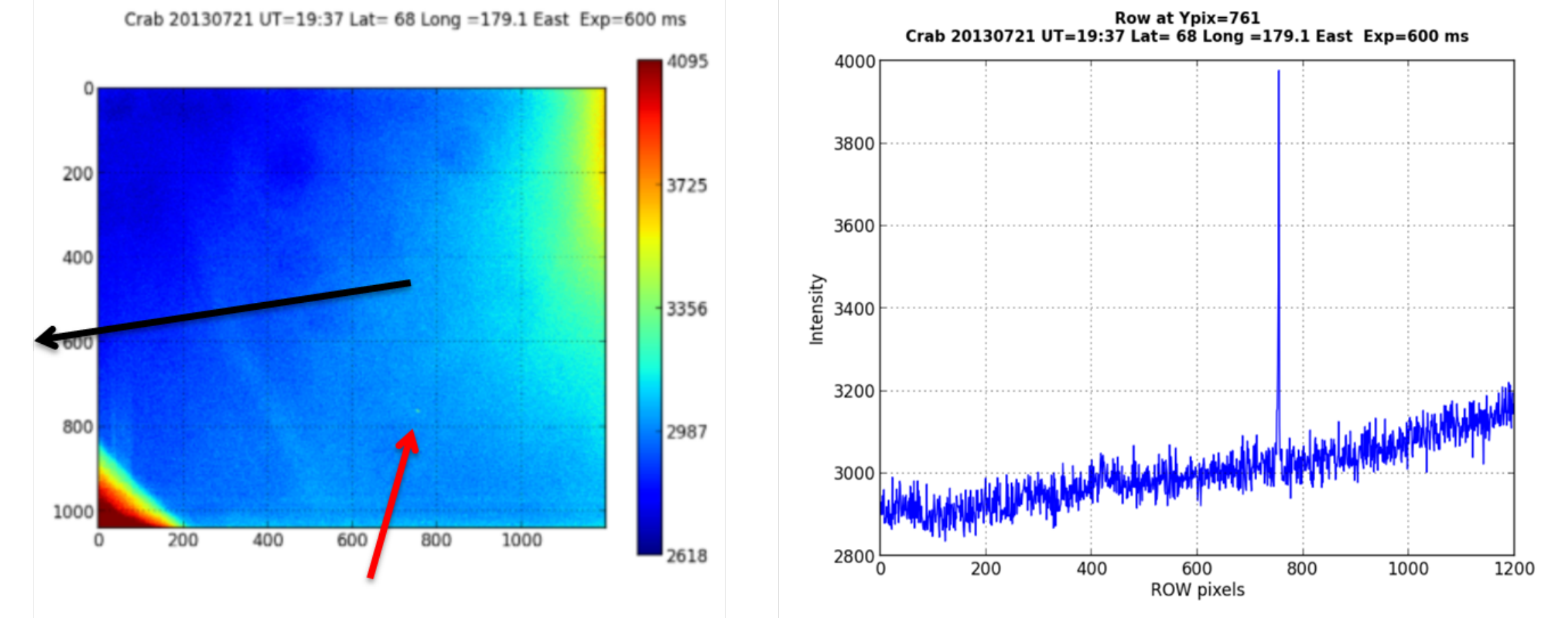}
\caption{\label{Star tracking Z Tau}STR star tracking during a Crab pointing on July 21st 2013, at 19:37~UT, with a \mbox{600 ms} exposure. 
The gondola position is \mbox{68.0$^\circ$ N}, \mbox{179.1$^\circ$ E}, with an altitude of  \mbox{36.5 km}. Left: pixel intensity map, 
heavily influenced by background and stray light. The direction towards the Sun has been indicated by a black arrow. The magnitude~3.0 
guide star \mbox{$\zeta$ Tauri}, shown by the red arrow, is barely visible. Right: horizontal pixel intensities (at pixel row~761) showing the 
gradient in the image. The three-point criterion allows the star to be clearly identified.}
\end{figure}

During observations of the Crab, star tracking is complicated by the relatively small angular distance to the Sun\footnote{Future flights are 
likely to use the Sun itself as a tracking target. Star trackers were originally foreseen for a one day long maiden flight of PoGOLite planned for 
August and were retained when a long duration flight was pursued instead.}.  
The performance of the star trackers depends on, e.g., image focus, gondola stability and exposure time. 
By co-adding images with short exposure time the gondola movement becomes negligible.
As shown in Figure~\ref{Co-added short exposure}, the image point-spread function is symmetric, indicating that gondola motion 
and not focus ultimately dictates image quality. This shows that the temperature-compensating mechanical structure 
shown in Figure~\ref{STR mechanical design} is working adequately.
An example of a representative \mbox{600 ms} exposure STR image and a pixel intensity scan are shown in Figure~\ref{Star tracking Z Tau} for a Crab 
observation at 36.5~km altitude on July 21st 2013, at 19:37~UT.  The pixel intensity map is dominated by background and 
stray light.  
A three-point criterion is used to identify stars in the CCD image. 
At least three adjacent pixel intensity values in both horizontal and vertical directions must 
exceed the noise level to be identified as a star, with the highest value required for the central pixel.
Despite the severe image gradient, this method allows a magnitude 3.0 guide star to be clearly identified.   

\begin{figure}[!hbt]
\centering
\includegraphics[width=.85\textwidth]{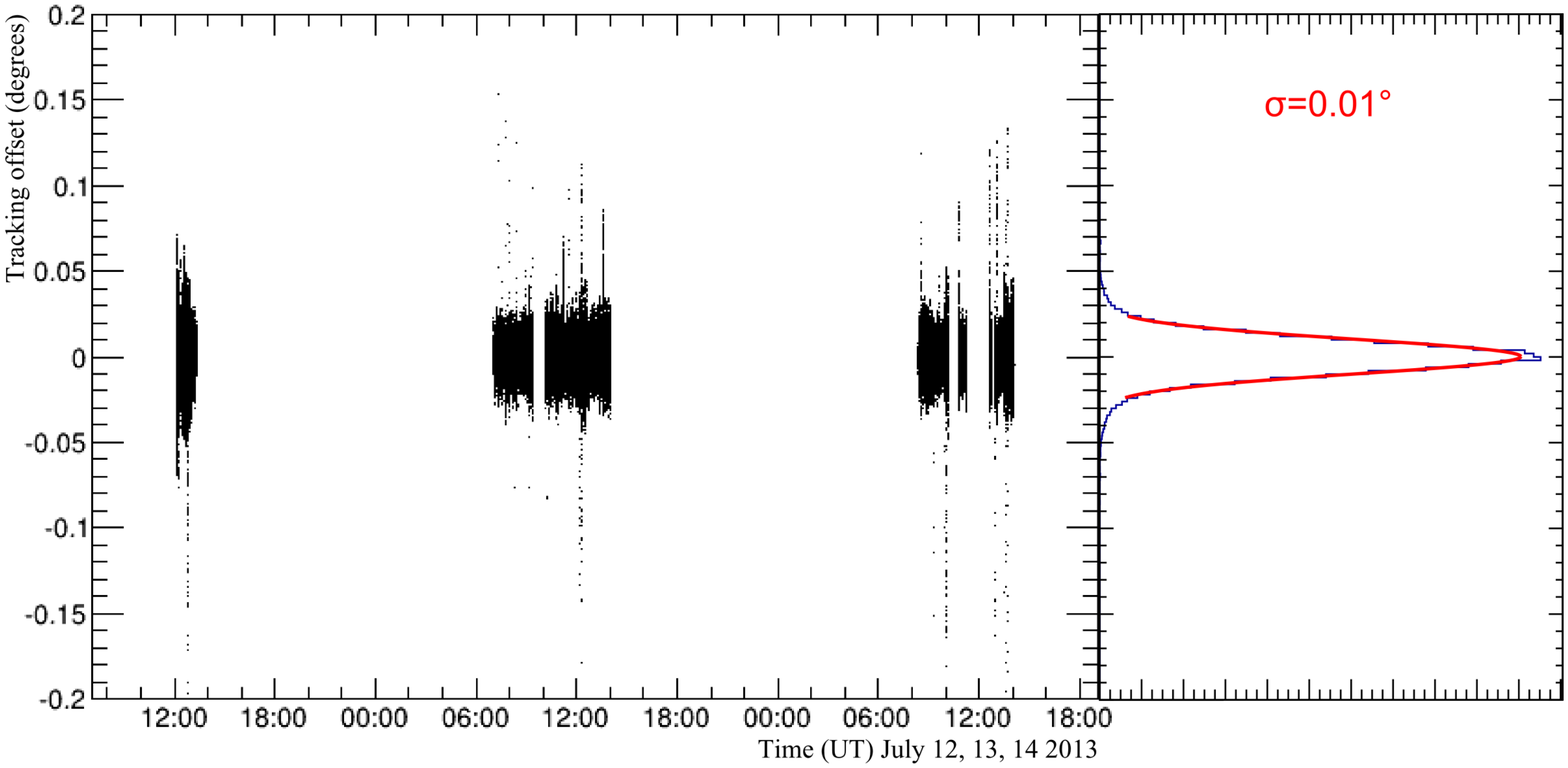}
\includegraphics[width=.85\textwidth]{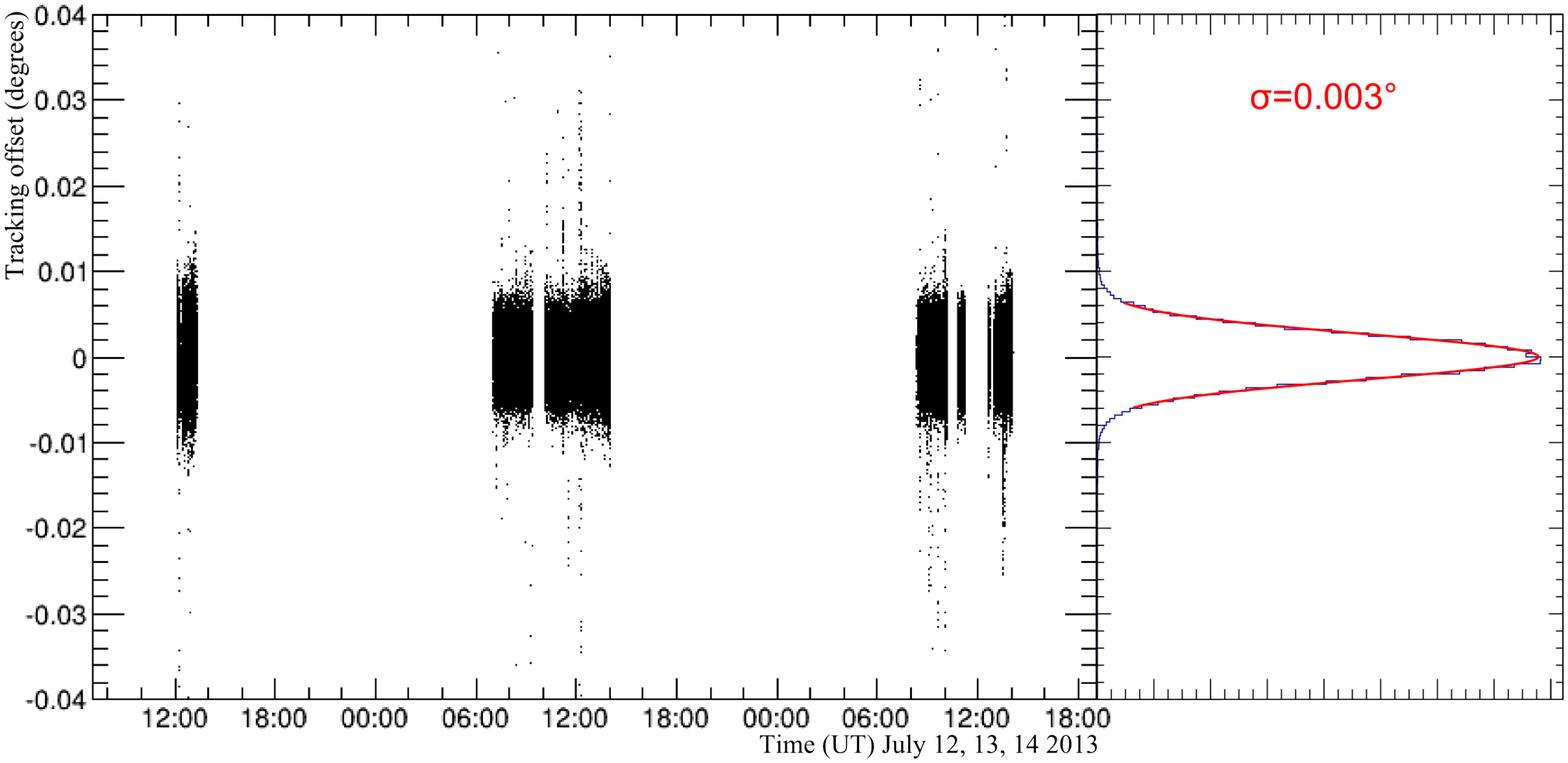}
\caption{\label{Pointing performance}{Angular distance between the set-point and reconstructed instrument pointing for azimuth (top) and 
elevation (bottom) for Crab observations on July 12th, 13th and 14th 2013. The left-hand side shows the offset as a function of time (UT) 
while the right-hand side is the projection.}}
\end{figure}

Although difficulties were experienced during star tracking, e.g. spurious reflections within the baffle assembly due to the proximity of the Sun to the star field or disturbances from passing satellites, the performance of the attitude control system was found to significantly exceed the design requirement.
The time dependent and integrated angular offset between the polarimeter pointing and the Crab set-point, is shown in azimuth and 
elevation in Figure~\ref{Pointing performance} for all Crab observations. The pointing accuracy is found to be an order of magnitude better 
than the design requirement  in both elevation and azimuth. 

%%%%%%%%%%%%%%%%%%%%%%%%%%%%%%%%%%%%%%%%%%%%%%%%%%%%%%%%%%%%%%%%%%%
\subsection{Observations}
\label{sec:obs}
\begin{figure}[!hbt]
\centering
\includegraphics[width=.85\textwidth]{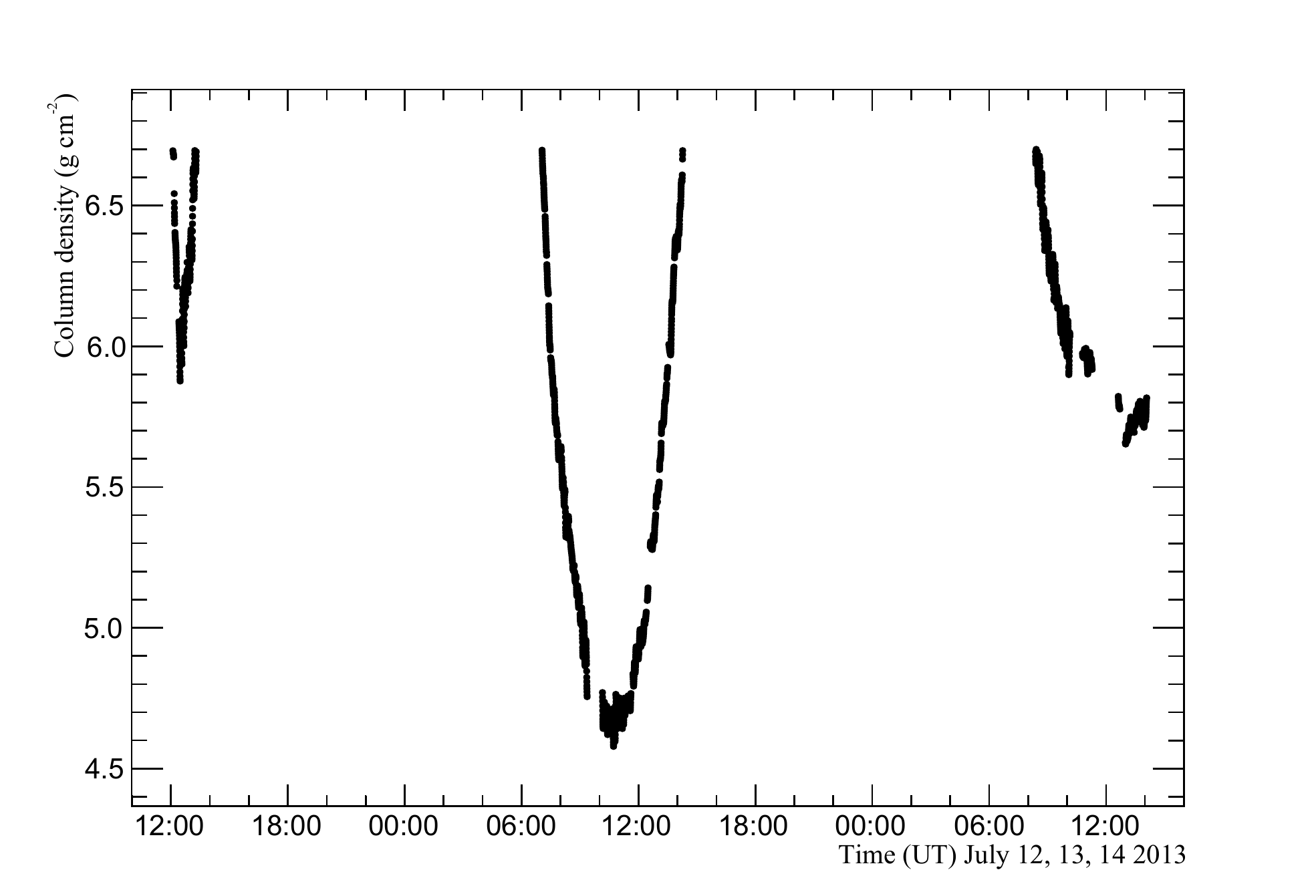}
\caption{\label{Crab observation column density}{Atmospheric column density during Crab observations determined from the measured 
atmospheric pressure (g/cm$^2$) and polarimeter elevation. The sharp transition in the left-most part of the curve is a result of the gondola still ascending 
during the first Crab observations. The Crab reaches a maximum elevation of $\sim$44$^\circ$. The altitude for the third Crab observation was 
anomalously low due to passage over the cool air above Greenland. }}
\end{figure}

\begin{figure}[!hbt]
\centering
\includegraphics[width=.85\textwidth]{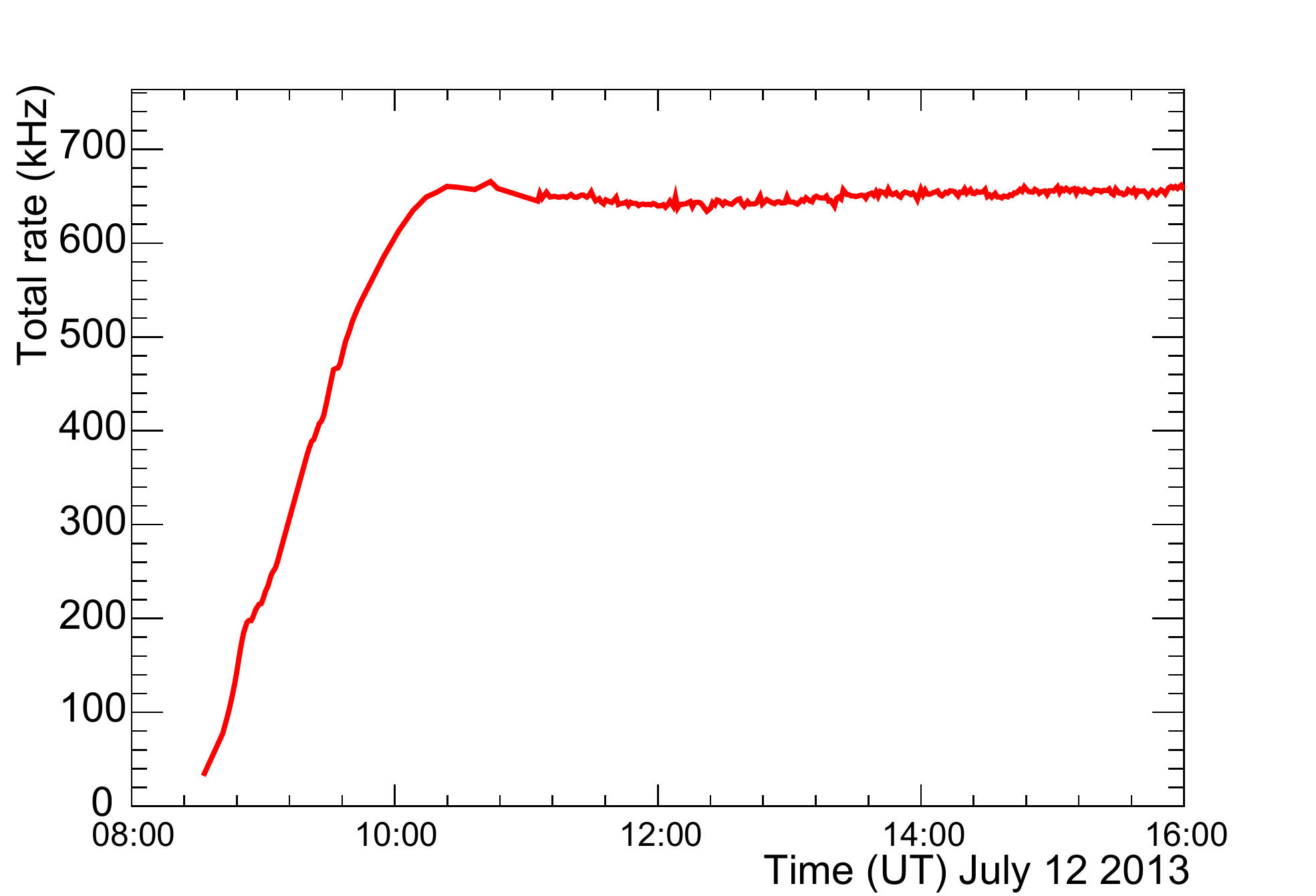}
\caption{\label{PDCAscent}{The evolution of the total PDC counting rate (histogram data) during ascent to float altitude. The hit threshold of 0.5~keV is applied. The maximum counting rate occurs at an altitude of 23 km at 10:00 UT. Float altitude, $\sim$40~km, is reached around 12:30 UT.}}
\end{figure}

Three Crab observations were made in the period July 12th-14th 2013 for a total observation time of 549 minutes with a reconstructed pointing direction 
within \mbox{0.1$^{\circ}$} of the Crab direction. The atmospheric column density during observations is shown in Figure~\ref{Crab observation column density}, yielding a time-averaged column density of 5.7 g/cm$^2$. At 30 keV (100 keV), the corresponding transmission probability is 13\% (41\%).
Figure~\ref{PDCAscent} illustrates the challenging background environment. The total PDC count rate (histogram data, $>$0.5~keV) rapidly increases during ascent to float altitude before stabilising at around 650~kHz. The corresponding total counting rate ($>$10~keV) for the side anticoincidence system is 300 kHz. The online veto system is used to reject background interactions, resulting in a manageable two-hit counting rate which forms the basis for polarisation measurements. 
After applying the online veto system (described in Section~\ref{Polarimeter electronics and control systems}), the two hit counting rate is 84~Hz for an expected Crab two hit rate of $\sim$2~Hz. 

Data reduction techniques developed for on-ground calibration tests~\cite{Ground calibration paper} needed to be complemented in order to achieve an acceptable signal-to-background performance for flight conditions. 
The temperature increase experienced by the waveform digitiser boards during Crab observations 
(see Section~\ref{sec:thermal}) affected the dynamic range of the charge-sensitive amplifiers depicted in Figure~\ref{DAQ overview}.
A temperature dependent correction was therefore applied to waveform data in order to recover the correct baseline level. Pre-trigger sample points could then be reliably used to reject events with variable baselines, e.g. due to preceding large signals or photomultiplier after-pulses.
Additionally, fake two-hit polarisation events including a scattering from a slow scintillator collimator to a fast scintillator in the same PDC resulted in super-imposed waveforms which were difficult to reject due to the relatively low amplitude signal from the slow scintillator. A Principal Component Analysis (PCA) method was developed to reject such events. 
Waveforms were characterised using ten variables, e.g. the fast and slow output values, the peak value of the waveform, the number of sample points in the waveform rise and the starting sample point of the rise. 
The PCA method allows this multivariate (ten dimensional) approach to be reduced to a pair (two dimensions) of uncorrelated components. Two dimensional selections were placed on the resulting principal and secondary components after performing a basis transformation. A standard selection on the fast branch, as described in Section~\ref{Polarimeter electronics and control systems}, was also applied. 

\begin{figure}[!hbt]
\centering
\includegraphics[width=.85\textwidth]{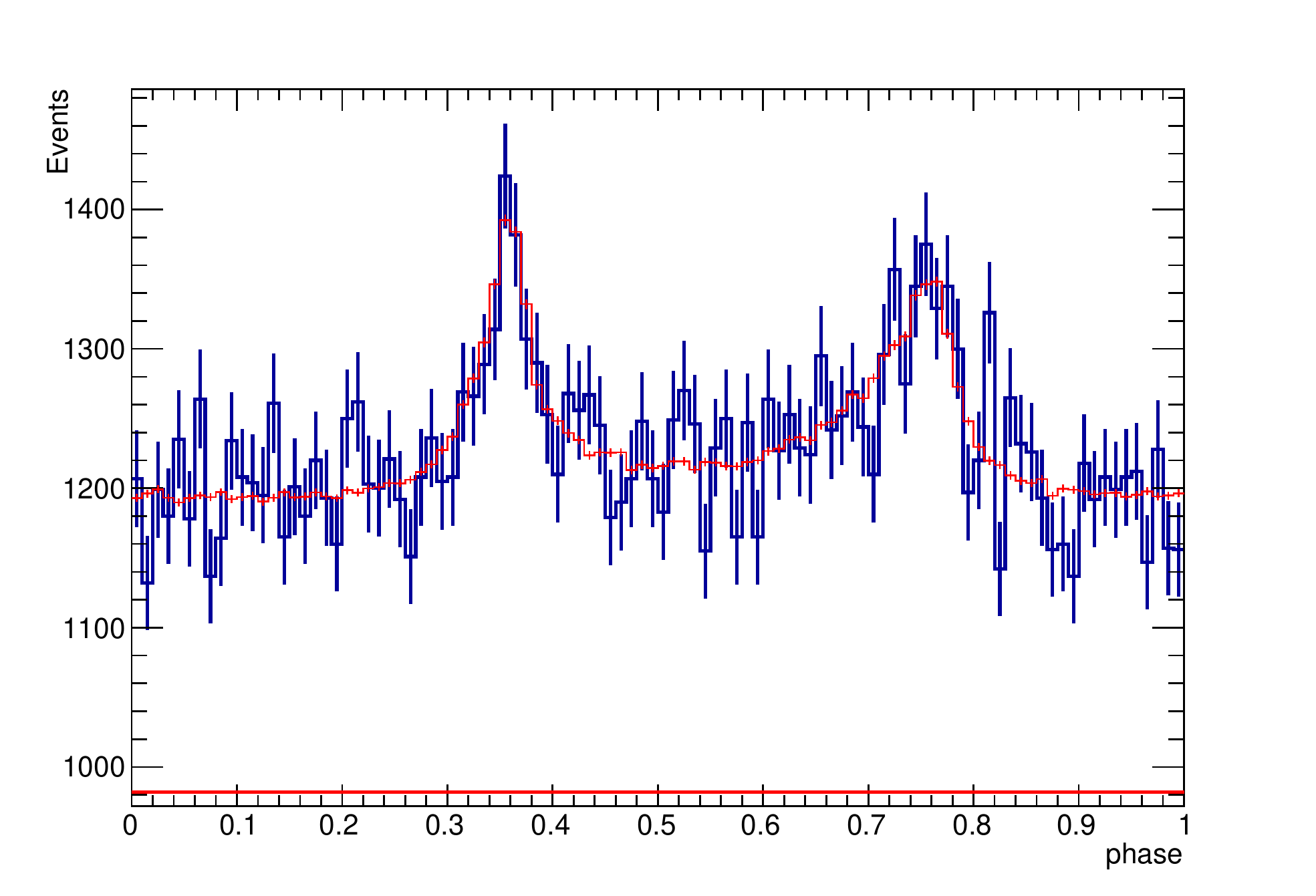}
\caption{\label{Crab light-curve}{The light-curve obtained for Crab observations (blue histogram) and a reference light-curve derived from 
Suzaku data (red histogram~\cite{Suzaku public data}). A $\chi^2$ optimisation for the signal-to-background 
ratio and normalisation is used to fit the Suzaku data, resulting in a derived signal-to-background ratio of 0.25 for the Crab observation.}}
\end{figure}

Selections in the resulting two-dimensional plane are iteratively evaluated by minimising an established figure-of-merit for X-ray polarimeters, 
the Minimum Detectable Polarisation (MDP)~\cite{MDP} which is sensitive to the signal-to-background ratio in observations.
The quality of the barycentred pulsar light-curve is also considered using the H-test statistic~\cite{de Jager 1 & 2}. Neither of these metrics are related to polarisation observables such as the modulation curve.
The H-test statistic is well suited to data with a low signal-to-background ratio and 
does not require a priori assumptions regarding the shape of the pulsed distribution. 
Periodicity was searched for in a narrow band around the known Crab pulsar spin period~\cite{Jodrell Bank}. For final selections, a H-test 
statistic of 1.5 ($P=0.6$) was determined for background observations, compared to 132 ($P=10^{-23}$) for all Crab observations, where $P$ is the probability that the observed periodicity would arise purely from statistical fluctuations.
The reconstructed light-curve is shown in Figure~\ref{Crab light-curve}. Polarimeter events with reconstructed energy within the range \mbox{25 - 110 keV} are used. This light-curve is compared to that derived from the Suzaku hard X-ray detector~\cite{Suzaku HXD} over a comparable energy 
range: \mbox{10 - 70 keV}. 
Since the signal-to-background ratio of the Suzaku measurement is known, a pure signal reference light-curve can be obtained. By fitting this reference light-curve, diluted by an appropriate fraction of background, the PoGOLite signal-to-background ratio is found to be 0.25$\pm$0.03. The Crab M$_{100}$ is determined as (21.4$\pm$1.5)\%, and the MDP for all Crab observations is (28.4$\pm$2.2)\%. 

\begin{figure}[!hbt]
\centering
\includegraphics[width=\textwidth]{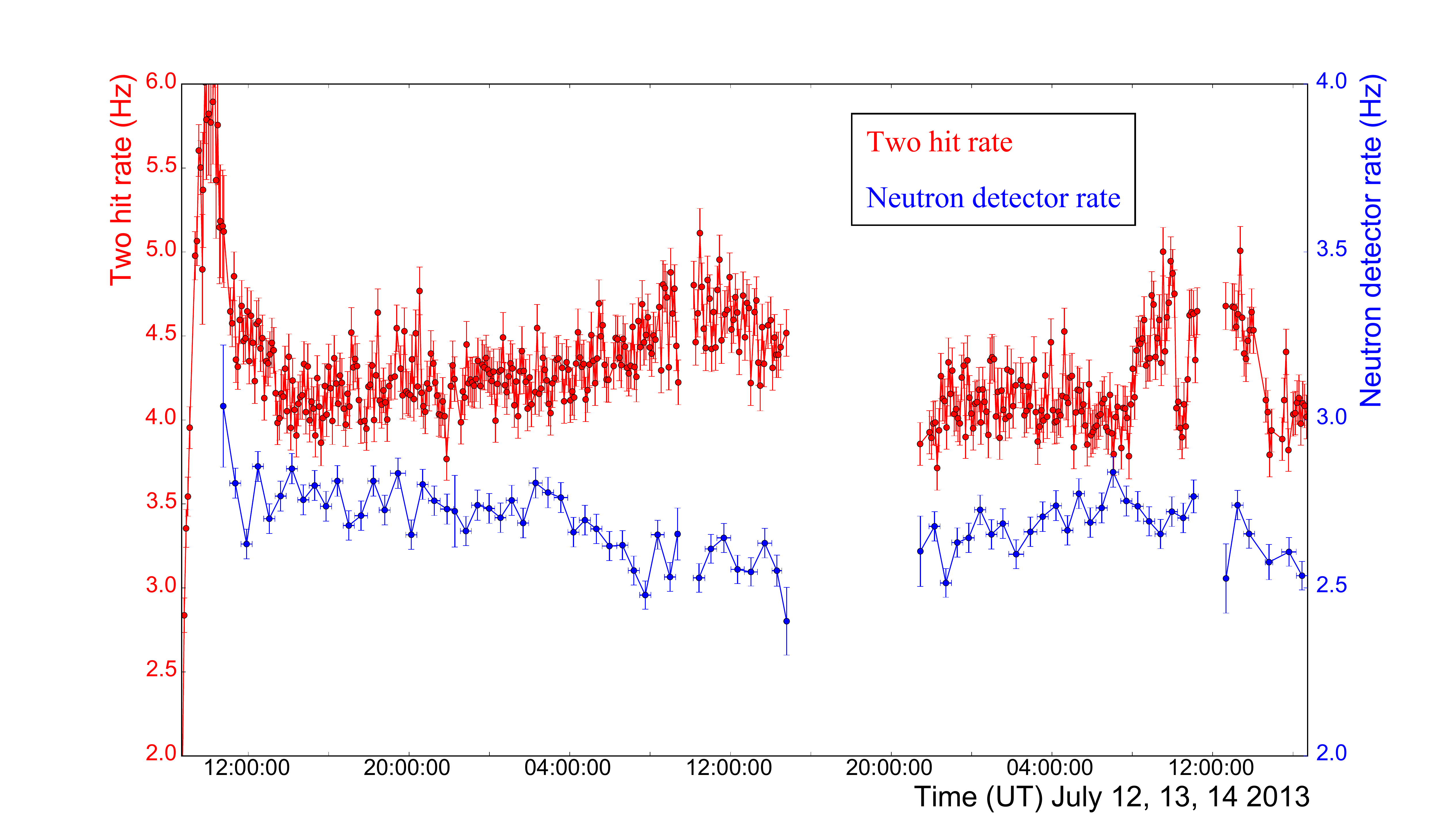}
\caption{\label{twohit}The time development of the two hit polarisation counting rate (red points, left vertical scale) and the neutron detector counting rate (blue points, right vertical scale). The two hit rate increases during Crab observations. The first Crab observation around 12:00 UT on July 12th is superimposed on a decreasing count rate trend as float altitude is reached. The decrease in count rate around 10:00 UT during the final observation is due to an off-source background measurement. The neutron count rate decrease significantly during the second Crab observation on July 13th.}
\end{figure}

The time evolution of the two hit rate is shown in Figure~\ref{twohit} and confirms the light-curve signal-to-background prediction. Data from the LiCAF-based neutron detector can be used to gauge relative changes in the dominant background due to atmospheric neutrons. The decrease in neutron background coincident with the second Crab observation (July 14th) is confirmed by a background model~\cite{MKo paper} based on the gondola altitude, longitude and solar activity estimated from ground-based neutron monitor data.
%

%%%%%%%%%%%%%%%%%%%%%%%%%%%%%%%%%%%%%%%%%%%%%%%%%%%%%%%%%%%%%%%%%%%
\section{Discussion and outlook}
\label{Discussion/summary/outlook}
The primary aim of the PoGOLite Pathfinder mission is to evaluate the experimental approach chosen to measure the polarisation of hard 
X-ray emissions from the Crab. During the $\sim$14 day long flight, three Crab observations were completed before the polarimeter control system failed. 
Observations were hampered by an inadequate thermal design of the polarimeter 
control electronics which significantly reduced the observation time. These issues are the focus of a post-flight engineering review.  A 
pointing precision in both azimuth and elevation of one order of magnitude better than the design goal was achieved. Signal from the Crab 
was unambiguously detected for a signal-to-background ratio of 0.25$\pm$0.03. In the PoGOLite energy band, the Crab polarisation fraction is likely to lie below the MDP.  A polarisation measurement is still possible but requires the development of a method to deal with the resulting non-Gaussianities in the determination of polarisation parameters resulting from measuring a positive-definite quantity close to the limit of observability. Such a method and a polarisation result will be presented in a future publication. 

A 5~day duration flight to 
Canada is used as a baseline for possible future flights. For the current polarimeter design, the expected Crab minimum detectable polarisation (MDP) for such a flight is $\sim
$20\%, using the flight conditions experienced in 2013 (e.g. altitude profile, signal-to-background performance). The 
observation time takes into account a 50\% duty cycle due to off-Crab (5$^\circ$ in azimuth) observations in order to measure 
background (not possible during the 2013 flight due to over-heating issues). 
Using data from the Pathfinder flight, a number of improvements to the polarimeter design have been identified in order to 
improve this performance.
\begin{itemize}
\item Post-flight calibration tests~\cite{Ground calibration paper} revealed unintentional leakage of scintillation light between PDCs. 
Although the magnitude of the effect is small, it degrades polarimetric response significantly, e.g. M$_{100}$ reduces by approximately a 
third. It is straight-forward to address this issue once the PDCs are removed from the polarimeter. The application of new reflective 
wrapping materials will also improve the PDC light-yield and further improve the performance, especially for low deposited energies. 
\item For the flight data, waveforms were found to be distorted due to low energy deposits in the slow scintillator which could not be 
separated from fast scintillator signals. The slow plastic scintillator collimator will therefore be replaced with a purely passive collimator 
comprising (from inside to outside) copper, tin and lead.   
\item The fast scattering scintillators will be shortened to 12~cm in length in order to reduce the target volume for background neutron 
interactions while not affecting efficiency for source X-ray detection. As a by-product, the light-yield will be improved.
\item The neutron shield will be further optimised, e.g. reducing gaps in the shielding during integration of the polarimeter. 
\end{itemize}

A re-flight of an upgraded polarimeter has been proposed for summer 2016 from Esrange in order to make polarimetric studies of the 
Crab (and Cygnus X-1, if possible) with a significant improvement in performance compared to the 2013 flight. For a 5 day flight (assuming 
2013 flight conditions) with multiple Crab exposures, the resulting MDP is significantly improved at 8.2\%. 
It is illustrative to consider the expected performance of the 'full-size' PoGOLite concept with 217 PDCs. For an equivalent flight, the MDP 
can be conservatively expected to improve by a factor $\sqrt{A_{217}/A_{61}}$, where $A_n$ is the collecting area. The resulting MDP 
prediction is 4.3\% for 1 Crab sources (15-20\% estimated for 200~mCrab sources). It is however noted that there are major 
engineering challenges to address before a larger polarimeter can be constructed. For example, the current gondola mass of 1850 kg is 
close to the limit for the 1.1 Mm$^3$ balloon used - the largest currently available.

%%%%%%%%%%%%%%%%%%%%%%%%%%%%%%%%%%%%%%%%%%%%%%%%%%%%%%%%%%%%%%%%%%
\begin{acknowledgements}

The PoGOLite Collaboration acknowledges funding received from The Swedish National Space Board, The Knut and Alice Wallenberg 
Foundation, 
The Swedish Research Council and the G\"{o}ran Gustafsson Foundation. The SSC Esrange Space Centre is thanked for their considerable 
support 
and expertise during the launch build-up and flight campaign of the PoGOLite Pathfinder. 
The government of the Russian Federation is thanked for permitting a circumpolar flight. 
Tim Thurston designed the PoGOLite mechanics which was primarily manufactured by the workshop at AlbaNova University Centre.
Gilles Bogaert provided VM2000 for the scintillator array. All past members of the PoGOLite Collaboration not listed as authors on this 
paper are thanked for their important contributions to the development of the project. 
\end{acknowledgements}


\end{document}